\documentclass[conference]{IEEEtran}
\usepackage{booktabs} 
\usepackage{caption}
\usepackage{subcaption}
\usepackage{placeins}
\usepackage{xcolor}
\usepackage{float}
\usepackage{url}
\usepackage{epstopdf}
\usepackage{graphics,graphicx}
\usepackage{array}
\usepackage{pifont}
\usepackage{multirow}
\usepackage{amssymb}
\usepackage{amsmath}
\usepackage{algorithm, algpseudocode}

\usepackage{paralist, tabularx}

\usepackage{color}

\begin{document}

\title{Fairness for Whom? Understanding the Reader's Perception of Fairness in Text Summarization}


\author{\IEEEauthorblockN{Anurag Shandilya, Abhisek Dash}
\IEEEauthorblockA{
Indian Institute of Technology Kharagpur, India
}
\and
\IEEEauthorblockN{Abhijnan Chakraborty}
\IEEEauthorblockA{Max Planck Institute for Software Systems, Germany
}
\and
\IEEEauthorblockN{Kripabandhu Ghosh}
\IEEEauthorblockA{
Indian Institute of Science Education and Research Kolkata, India
}

\and
\IEEEauthorblockN{Saptarshi Ghosh}
\IEEEauthorblockA{
Indian Institute of Technology Kharagpur, India
}
}

\maketitle

\begin{abstract}
With the surge in user-generated textual information, there has been a recent increase in the use of summarization algorithms for providing an overview of the extensive content. 
Traditional metrics for evaluation of these algorithms (e.g. ROUGE scores) rely on matching algorithmic summaries to human-generated ones. However, it has been shown that when the textual contents are heterogeneous, e.g., when they come from different socially salient groups, most existing summarization algorithms represent the social groups very differently compared to their distribution in the original data. To mitigate such adverse impacts, some fairness-preserving summarization algorithms have also been proposed. 
All of these studies have considered normative notions of fairness from the perspective of writers of the contents, neglecting the readers' perceptions of the underlying fairness notions. To bridge this gap, in this work, we study the interplay between the fairness notions and how readers perceive them in textual summaries. 
Through our experiments, we show that reader's perception of fairness is often context-sensitive. Moreover, standard ROUGE evaluation metrics are unable to quantify the perceived (un)fairness of the summaries. 
To this end, we propose a human-in-the-loop metric and an automated graph-based methodology to quantify the perceived bias in textual summaries. We demonstrate their utility by quantifying the (un)fairness of several summaries of heterogeneous socio-political microblog datasets.\footnote{\textcolor{red}{This work has been accepted at International Workshop on Fair and Interpretable Learning Algorithms 2020 (FILA 2020), which was held in conjunction with IEEE BigData 2020. Please cite the version appearing in the proceedings.}}
\end{abstract}


%
\IEEEpeerreviewmaketitle

\section{Introduction}
\begin{figure}[t]
	\centering
	\begin{subfigure}{0.9\columnwidth}
		\includegraphics[width= \textwidth, height=4cm]{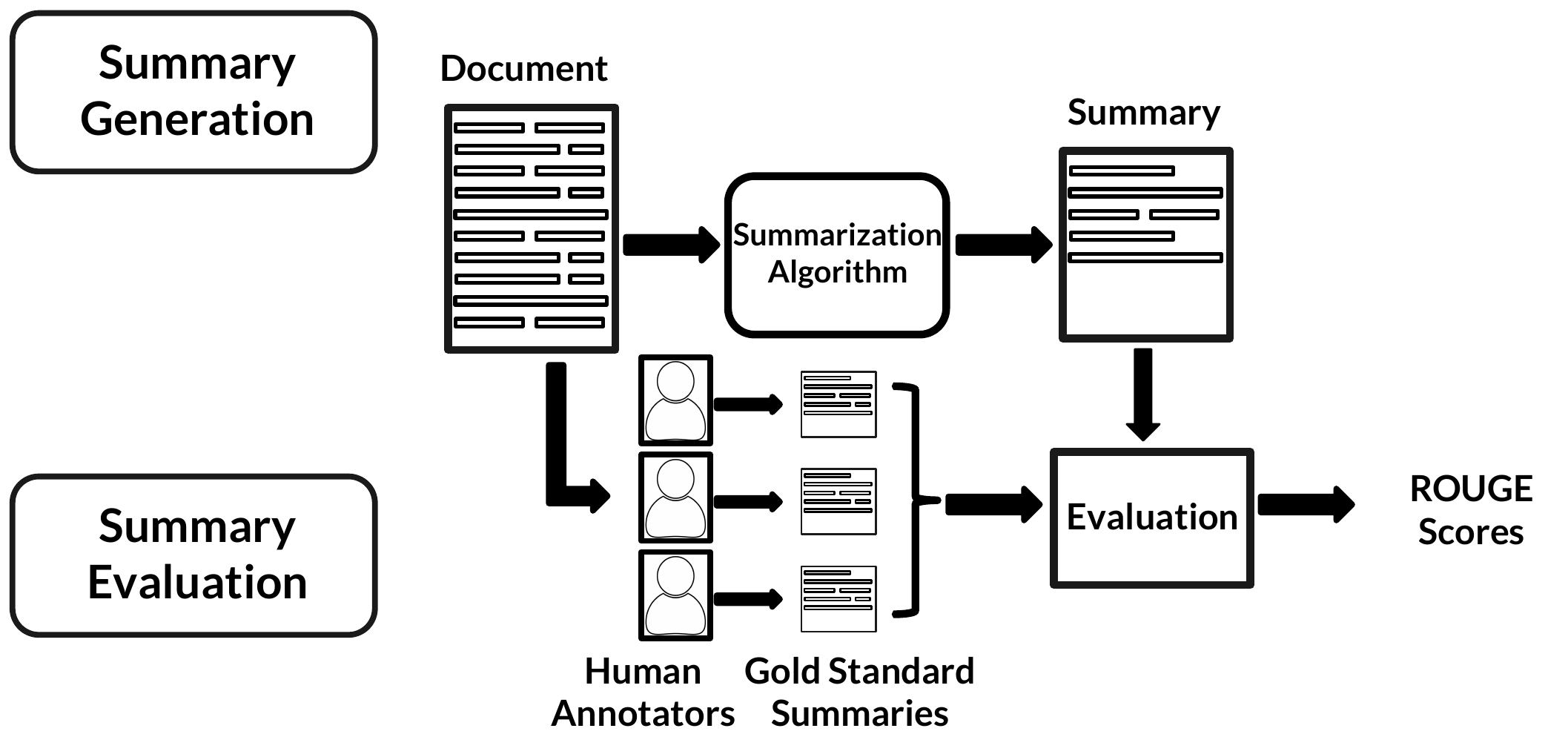}
	\end{subfigure}%
	\caption{{\bf A generic block diagram explaining text-summarization pipeline. Machine generated summaries are evaluated based on how well they match human written reference summaries. Metrics such as ROUGE scores quantify the goodness of such  automated summaries.}}
	\label{Fig: Summarization}
	\vspace{-6 mm}
\end{figure}

\noindent
With the surge in the amount of textual information on the Web, text summarization algorithms~\cite{text-summarization-survey,text-summarization-survey-2020} are increasingly being used to get a quick overview of the information. The standard framework for text summarization can be broadly divided into two parts: summary generation and summary evaluation (as shown in Figure~\ref{Fig: Summarization}). 
In summary generation, given a document or sometimes a set of documents, a summarization algorithm summarizes it. Generally, two kinds of summarization approaches are followed in the literature~\cite{text-summarization-survey,text-summarization-survey-2020} -- 
(i)~\textit{extractive summarization}, where the algorithms select sentences from the document to include in the summary, and 
(ii)~\textit{abstractive summarization:} where the algorithms produce natural language summaries. 

Traditional summarization algorithms are meant for summarizing homogeneous documents (e.g. 
news article(s) on a topic, or research paper(s)) and they have only focused on {\it summary-worthiness} of textual units while deciding on whether to include or exclude them in the summary. However, with the  growing popularity of social media websites, e.g. Facebook, Twitter, user-generated content constitutes a large chunk of the textual information generated on the Web today. On social media, different user groups discuss different socio-political issues, and it has been observed that they often have very different opinions on the same topic or event~\cite{dash2019summarizing,mukherjee2020read}. 
Hence, the textual information to be summarised has gradually become heterogeneous. 
In our prior work~\cite{dash2019summarizing}, we have shown that such text often contains very different opinions from people of different ideologies, social groups, etc. In many downstream applications, algorithm-generated summaries are consumed by people and hence they often play a vital role in shaping their opinion in different socio-political issues. Hence, along with summary quality, the fairness aspect of algorithmic summaries (that are produced by automatic summarization algorithms) have also become essential~\cite{shandilya2018fairness, dash2019summarizing}. Lately, this has led to different fair summarization algorithms for heterogeneous user generated textual units~\cite{dash2019summarizing, mukherjee2020read}.

\vspace{1 mm}
\noindent
\textbf{Evaluation of algorithmic summaries:} 
Traditionally, the evaluation of algorithmic summaries are carried out by evaluating how closely they match human-generated summaries.
The same source document (or set of documents) is given to a number of human annotators to summarize. Metrics like ROUGE~\cite{lin2004rouge}, ROUGE2.0~\cite{ganesan2015rouge}
are used to quantify the goodness of the algorithmic summaries. Even though these measures perform very well in evaluating the goodness of summaries (based on textual quality and readability etc.), they do not explicitly quantify the (un)fairness of an algorithmic summary. 
Moreover, this process of evaluation is often laborious and hence an expensive task. 
Evaluation in multi-document summarization is particularly expensive. It is reported that 3,000 hours of human effort is required to evaluate the summaries from the Document Understanding Conferences (DUC)~\cite{lin2004rouge}.  

\vspace{1 mm}
\noindent
\textbf{Drawbacks in the existing framework:} 
The existing fair summarization algorithms have mostly tried to incorporate normative representational fairness goals from the perspective of the content producers/writers in the final summary. 
However, {\it whether the summaries are perceived to be fair by the consumers/readers} is still up for debate. 
Additionally, the different existing approaches of evaluating summaries (the most popular being computation of ROUGE scores) have several limitations when it comes to quantification of fairness aspect of the summaries of heterogenous user-generated text corpora. 

\vspace{1 mm}
\noindent
\textbf{Current work:} In this work, we posit that in the context of summarization, fairness is highly context-dependent, and ideally involves multiple stakeholders. 
The most important stakeholders in a summarization set up are: producers or writers of the textual units, and consumers or readers of the final summary.\footnote{We use the word-pairs `producers' and `writers', as well as `consumers' and `readers' interchangeably throughout this paper.} 
However, the interpretation of fairness may vary when we envisage it from the reader's perspective. To this end, in this work, we investigate the interplay between the earlier proposed definitions of fairness in summarization and the consumers' perceptions of fairness, and how this interplay varies with the context of the underlying topic. Further, we also investigate the effectiveness of existing measures, e.g. ROUGE in quantifying the (un)fairness of a summary. 

Specifically, we seek for the answer to the following research questions (RQs). 
RQ1:~Is the readers' perception of (un)fairness in summaries context-dependent?, 
RQ2:~Do traditional metrics for summary quality such as ROUGE scores capture readers' perception of fairness of summaries?, and finally, 
RQ3:~Can a metric based on `representation of opinions’ better capture readers’ perception of (un)fairness in summaries? 
To answer the aforementioned RQs, we conducted a series of surveys on two socio-political datasets (obtained from~\cite{dash2019summarizing}) of microblogs/tweets related to (i)~the US Presidential Elections, and (ii)~the MeToo movement. 
Through the different analyses, the main contributions/observations of the present work can be summarised as follows:
\begin{enumerate}
	\item We show that readers can differentiate between fair and unfair summaries. However, the reasons why a summary is perceived to be (un)fair is context-dependent. In some cases, the perceived fairness agrees with standard representational fairness notions for demographic groups of producers; while in other cases, the perceived fairness seems to agree more with how fairly various opinions are represented in the summary.
	\item In either case, standard ROUGE metrics cannot capture the bias in summaries as perceived by the consumers.
	\item We propose a metric for perceived bias in a summary, based on manual identification of opinions in the input text, and then judging how well various opinions are represented in the summary.
	\item Finally, we propose a graph-based methodology for automatically measuring the bias in a summary. We observe that correlates well with the perceived opinion bias metric stated above.
\end{enumerate}

\section{Background and Related Work}
\noindent
In this section, we 
discuss a few relevant prior works on fairness in text summarization and motivate the present work by contextualizing it in the existing literature.  

\subsection{Fairness in text summarization}
\noindent
Much like in the fairness in ML literature~\cite{ali2019fairness,friedler2019comparative,patro2020fairrec}, the proposed methodologies for fair text summarization can be divided into three categories e.g., (1)~pre-processing, (2)~in-processing, (3)~post-processing based algorithms based on the stage at which fairness intervention is performed. 
In the pre-processing based algorithms, the dataset is fed to the summarization algorithms in a way such that the generated summaries will end up being fair. Similarly, in post processing algorithm, fairness interventions are applied on the output of standard summarization algorithms to generate fair summaries. 
Finally, in the in-processing based approach~\cite{dash2019summarizing, mukherjee2020read}, the algorithm designers often treat summarization as an optimization problem and solve the same by either modifying the optimization function or adding fairness constraints to generate fair summaries. Next, we briefly discuss the FairSumm algorithm that was proposed in~\cite{dash2019summarizing}.

\vspace{2 mm}
\noindent \textbf{FairSumm algorithm for fair summarization:} Our prior work~\cite{dash2019summarizing} developed an in-processing fair summarization algorithm, called `\textbf{FairSumm}'. 
FairSumm treats the summarization task as a sub-modular optimization problem with fairness constraints and solves it to maximize coverage and diversity across the textual units while adhering to standard fairness notions~\cite{dash2019summarizing}. Given a heterogeneous set of micro-blogs (coming from different socially salient groups), and a desired target representation of the groups, the algorithm produces extractive summaries that reconcile between textual quality of the summaries (as quantified by ROUGE scores), and fair representation of different social salient groups in the summary. 
For instance, FairSumm can be applied over a set of tweets posted by male and female authors, to obtain a good summary having equal fractions of tweets posted by male authors and tweets posted by female authors.
We shall be using \textit{FairSumm} algorithm extensively for the experiments throughout this paper.


\subsection{Notions of fairness in text summarization} \label{sub:fairness-notions}
\noindent
Most of the prior works on fair summarization deal with the idea of group fairness. Specifically, when the input data (e.g. tweets or reviews) are generated by users from different socially salient groups, the algorithms explicitly enforce the summaries to fairly represent these different groups. 

\vspace{2mm}
\noindent
\textbf{Equal Representation:} The notion of equality finds its roots in the field of morality and justice, which advocates for the redress of undeserved inequalities (e.g. inequalities of birth or due to natural endowment)\cite{rawls2009theory}. In the context of summarization this ensures that the final summary must include equal number of textual units coming from different socially salient groups.

\vspace{2mm}
\noindent
\textbf{Proportional Representation:} Often it may not be possible to equally  represent different user groups in the summary, especially if the input data contains very different proportions from different groups. 
Hence, we consider another notion of fairness: {\it Proportional Representation} (also known as \textit{Statistical Parity}~\cite{luong2011k}).
In the context of summarization, Proportional Representation requires that the proportion of content from different user groups in the summary should be  same as in the original input. 


\noindent These notions of fairness ensure that the probability of selecting an item is \textbf{independent} of which user group generated it.

\subsection{Drawbacks in the current literature}
\noindent
The process of summarizing involves two parties: namely producers of the information a.k.a `writers' and consumers of summarized information a.k.a `readers'. All of the prior works on fairness in summarization have attempted to ensure the fair representation of the producers; whereas the fairness toward consumers or readers has been completely ignored. 
The inclusion or exclusion of certain opinions/voices tend to have the maximum effect on the consumers of the summaries. As the summary is what is read by the consumers, the summary shapes their opinion on the topic. Hence bias in the final summary can have severe impact on shaping the public discourse. Hence, in this work we focus on exploring the interplay of existing fairness definitions and how they are perceived by the readers.

\vspace{2mm}
\noindent
\textbf{Limitations of existing measures in quantification of (un)fairness in summaries:} For evaluation of algorithm-generated summaries, all of the prior works have evaluated the generated summaries based on ROUGE metric. However, in this work, we observe that ROUGE metric is unable to capture the (un)fairness aspect of the generated summaries. 
To this end, in this work we also propose a metric for perceived fairness of textual summaries. Further, we also propose an automated quantification of the perceived bias of textual summaries that correlates significantly with the aforementioned perceived fairness. 

To the best of our knowledge, this is the first work towards quantification of (un)fairness in summaries, and understanding the interplay between the perceived fairness in text summarization from the perspective of both writers and readers of the textual content.

\section{Datasets} 
\label{sec:datasets}
\noindent We reuse the following two datasets from our prior work~\cite{dash2019summarizing}.

\vspace{2mm}
\noindent \textbf{(1) US-Election dataset:} This dataset, originally provided by Darwish et al.\cite{darwish2017trump}, contains English tweets
posted during the 2016 US Presidential election. Each tweet is annotated as supporting or attacking
one of the presidential candidates (Donald Trump and Hillary Clinton) or neutral or attacking both.
For simplicity, we grouped the tweets into three classes: 
(i)~Pro-Republican: tweets which support
Trump and / or attack Clinton, 
(ii)~Pro-Democratic: tweets which support Clinton and / or attack Trump, and 
(iii)~Neutral: tweets which are neutral or attack both candidates. 


\vspace{2mm}
\noindent \textbf{(2) MeToo dataset:} We collected a set of tweets related to the MeToo movement in October 2018.
Specifically, we collected English tweets containing the hashtag `\#MeToo' using the Twitter Search API. 
We asked three human annotators to examine the name and bio of the Twitter accounts who posted the tweets.
The annotators observed three classes of tweets based on who posted the tweets -- (i)~tweets posted by male users, 
(ii)~tweets posted by female users, and
(iii)~tweets posted by organizations (mainly news media agencies). 
Also, there were many tweets for which the annotators could not understand the
type/gender of the user posting the tweet. For purpose of this study, we decided to focus only on
those tweets for which all the annotators were certain that they were written by male users or female users. 

\vspace{2mm}
\noindent From each of these two datasets, we selected a set of $50$ tweets, having an equal representation of the different demographic groups. In other words, we selected $50$ tweets from the USElection dataset, containing $17$ pro-Democratic tweets, $17$ pro-Republican tweets, and $16$ neutral tweets.
Similarly, we selected $50$ tweets from the MeToo dataset, containing $25$ tweets posted by male users and $25$ tweets posted by female users. 
While selecting these two sets of $50$ tweets, we ensured choosing distinct tweets  (for which we removed near-duplicates) that were well-formed and informative.
All experiments in this paper are conducted over these two sets of $50$ tweets each.

In the rest of this paper, we conduct a number of surveys and experiments on the aforementioned datasets in pursuit of answers to the RQs mentioned in the introduction. 

\section{Understanding Consumers' Perception of Fairness in Summaries}
\label{sec:consumer-perception}
\noindent
In this section, we investigate the  \textbf{RQ$1$} stated in the introduction-- whether readers' perception of (un)fairness in summaries is context dependent. 
To this end, we first generate summaries having different levels of biases, and then conduct a survey to understand how consumers (human annotators) perceive the bias/fairness of these summaries.

\subsection{Generating differently biased summaries}

\noindent
We consider a set of $50$ tweets from the US elections dataset ($17$ pro-Democratic tweets, $17$ pro-Republican tweets and $16$ neutral tweets), which are not repetitive in nature. We apply the FairSumm algorithm~\cite{dash2019summarizing} on this set of tweets to generate summaries of length $15$ tweets, having a wide variety of bias (from completely biased towards pro-Republican ideology to completely biased towards pro-Democratic ideology). 
To this end, we fix a certain number of neutral tweets, and then vary the number of pro-Republican and pro-Democratic tweets to create variously biased summaries.
Specifically, we create two batches of summaries, one batch with 3 neutral tweets each, and the other batch with 5 neutral tweets each. 

\vspace{2mm}
The first batch of summaries with 3 neutral tweets each, which we term as {\bf FairSumm-US-Batch1}, contains the following summaries (each of length $15$ tweets):
\begin{enumerate}
\item 00 pro-Rep tweets, 12 pro-Dem tweets, 03 neutral tweets -- actually very unfair summary
\item 02 pro-Rep tweets, 10 pro-Dem tweets, 03 neutral tweets -- actually very unfair summary
\item 04 pro-Rep tweets, 08 pro-Dem tweets, 03 neutral tweets
\item 06 pro-Rep tweets, 06 pro-Dem tweets, 03 neutral tweets -- actually very fair summary
\item 08 pro-Rep tweets, 04 pro-Dem tweets, 03 neutral tweets
\item 10 pro-Rep tweets, 02 pro-Dem tweets, 03 neutral tweets -- actually very unfair summary
\item 12 pro-Rep tweets, 00 pro-Dem tweets, 03 neutral tweets -- actually very unfair summary
\end{enumerate}

\vspace{2mm}
The second batch of summaries with 5 neutral tweets each, which we term as {\bf FairSumm-US-Batch2}, contains the following summaries (each of length $15$ tweets):
\begin{enumerate}
\item 00 pro-Rep tweets, 10 pro-Dem tweets, 05 neutral tweets -- actually very unfair summary
\item 02 pro-Rep tweets, 08 pro-Dem tweets, 05 neutral tweets -- actually very unfair summary
\item 04 pro-Rep tweets, 06 pro-Dem tweets, 05 neutral tweets -- actually very fair summary
\item 06 pro-Rep tweets, 04 pro-Dem tweets, 05 neutral tweets -- actually very fair summary
\item 08 pro-Rep tweets, 02 pro-Dem tweets, 05 neutral tweets -- actually very unfair summary
\item 10 pro-Rep tweets, 00 pro-Dem tweets, 05 neutral tweets -- actually very unfair summary
\end{enumerate}

\vspace{2mm}
Similarly we consider a set of $50$ tweets from the MeToo dataset containing $25$ tweets posted by male users and $25$ tweets posted by female users (as stated in Section~\ref{sec:datasets}).
We then apply FairSumm to generate the following summaries of length $15$ tweets each, having a wide variation of bias (from completely biased towards tweets posted by male users to completely biased towards tweets posted by female users).
We call this batch of summaries {\bf FairSumm-MeToo}, which contains the following summaries (each of length $15$ tweets):
\begin{enumerate}
\item 00 Male tweets, 15 Female tweets -- actually very unfair summary
\item 02 Male tweets, 13 Female tweets -- actually very unfair summary
\item 04 Male tweets, 11 Female tweets
\item 06 Male tweets, 09 Female tweets
\item 08 Male tweets, 07 Female tweets -- actually very fair summary
\item 10 Male tweets, 05 Female tweets
\item 12 Male tweets, 03 Female tweets -- actually very unfair summary
\item 14 Male tweets, 01 Female tweets -- actually very unfair summary
\end{enumerate}

\noindent It can be noted that, for all these summaries generated using the FairSumm algorithm, the actual biases are known in terms of the number of tweets included in a summary from the different perspectives. 
We will next check how the bias/fairness of these summaries is viewed by consumers (human annotators).

\begin{table*}
\center
\begin{tabular}{|p{0.95\textwidth}|}
\hline  
Hillary has derogatory titles for anyone not voting for her.\textbackslash
Unlike Hillary, Trump is facing rape charges.\textbackslash Nothing  will deter Trump and he will not stop fighting for you.\textbackslash
Bill Clinton has admitted that Obamacare is bad and Hillary is pissed about it.\textbackslash
Donald Trump claims credit for terrorist acts, just like terrorists.\textbackslash
Hillary is  only one that can make college affordable.\textbackslash
Trump says he has come on top in the Presidential debate.\textbackslash
17 out of 20 people feel that Hillary is winning.\textbackslash
Trump claims that sources that report negatively about his campaign are not to be trusted.\textbackslash
We know the net worth of Hillary, cause she has disclosed her assets.\textbackslash
Hillary thinks she has a solid strategy to defeat ISIS while Trump has none.\textbackslash
Some Trump supporters want him to win so that they can abuse women they want.\textbackslash
Iraq is getting ready for a battle to reclaim Mosul. \\

\hline \hline
Hillary shames everyone and thinks anyone not voting for her is stupid.\textbackslash 
Trump thinks Hillary  is crooked.\textbackslash
 Trump refuses to accept that the current POTUS was born in America \textbackslash
Obamacare is bad and Hillary is not happy with what Bill Clinton said about it.\textbackslash
Hillary doesn’t have the drive to make America great again.\textbackslash
People who are cancelling subscriptions to Dallas and Arizona newspapers are smart.\textbackslash
For people who don’t wanna vote, they need to be told that only Hillary can get rid of their huge college debt.\textbackslash
Trump thinks Hilary has been fighting ISIS without success for years and now it’s time for a change.\textbackslash
Trump thinks Hillary has told lies throughout her life and has sold America’s interests.\textbackslash
The way Hillary is handling the e-mail case, she is unfit for the post of President.\textbackslash
Hillary is a proponent of more love and kindness in America.\textbackslash
Hillary has a solid strategy to defeat ISIS unlike Trump.\textbackslash
Some guys want Trump to win so that they can oppress women.\textbackslash
We should be thankful to every nation that helped bring Paris agreement into action.\textbackslash
Shooting of unarmed Black men is unacceptable.\textbackslash
Every women in this country deserves to be free from harm and fear.\textbackslash
Charlotte should release police video of the Keith Lamont Scott shooting without delay. \\
 
\hline 
\end{tabular}
\caption{\textbf{Set of distinct opinions (separated by \textbackslash) identified by two of the annotators, from the set of tweets related to US Elections.}}
\vspace{-5mm}
\label{tab:opinions}
\end{table*}

\subsection{Understanding consumers' perception of (un)fairness}
\noindent
We start with a group of six annotators (3 males and 3 females) who have substantial knowledge of US politics and the MeToo phenomenon, and are in the age group of 18--30 years. 
We used a questionnaire to ascertain their knowledge of US politics and the MeToo movement. Also the annotators are familiar with use of social media platforms including Twitter, and none of the annotators is an author of this paper.

The annotators were first asked to go over the two sets of $50$ tweets each (one on USElection, and the other on MeToo) and to note down every distinct {\it opinion} expressed in the tweets.
An {\it opinion} is defined as a unique idea/information being conveyed by a tweet, hitherto not covered by any other/previous tweet.
Note that the annotators were only shown the text of the tweets; they were {\it not} told anything about the gender / political ideology of the users who authored the tweets. 
There was no limit to how many opinions they may identify, however each opinion was required to be confined to a maximum of two sentences. 
Table~\ref{tab:opinions} tabulates the opinions identified by two of the six annotators, from the set of $50$ tweets related to the US Elections.

Thereafter, the annotators are shown the summaries from the FairSumm-US-Batch1, FairSumm-US-Batch2 and FairSumm-MeToo batches in random order. 
They were asked to judge the fairness of each summary, and label each summary with one of the following labels:
\begin{enumerate}
    \item Very Fair Representation
    \item Somewhat Fair Representation
    \item Somewhat Unfair Representation
    \item Very Unfair Representation
\end{enumerate}
Along with labeling each summary, they were also asked to provide a reasoning for their judgement. In other words, they were asked to indicate the factor(s) based on which they were judging a summary to be fair/unfair:
\begin{enumerate}
    \item Fair/unfair representation of political/gender groups
    \item Fair/Unfair representation of political/contextual opinions
    \item Fair/Unfair representation of both: political/gender groups and political/contextual opinions
    \item Any other reason (requires a subjective response)
\end{enumerate}


Now we examine how the consumer’s perception of fairness varies across different contexts/scenarios. 
To this end, we plot the fraction of annotators who have annotated a summary as either `very fair representation' of the input text, or as `very unfair representation' of the input text. 
These two fractions are termed as `very fair approval fraction' and `very unfair approval fraction' respectively. 

\begin{figure*}[t]
	\centering
	\begin{subfigure}{0.65\columnwidth}
		\centering
		\includegraphics[width=\textwidth, height=3cm]{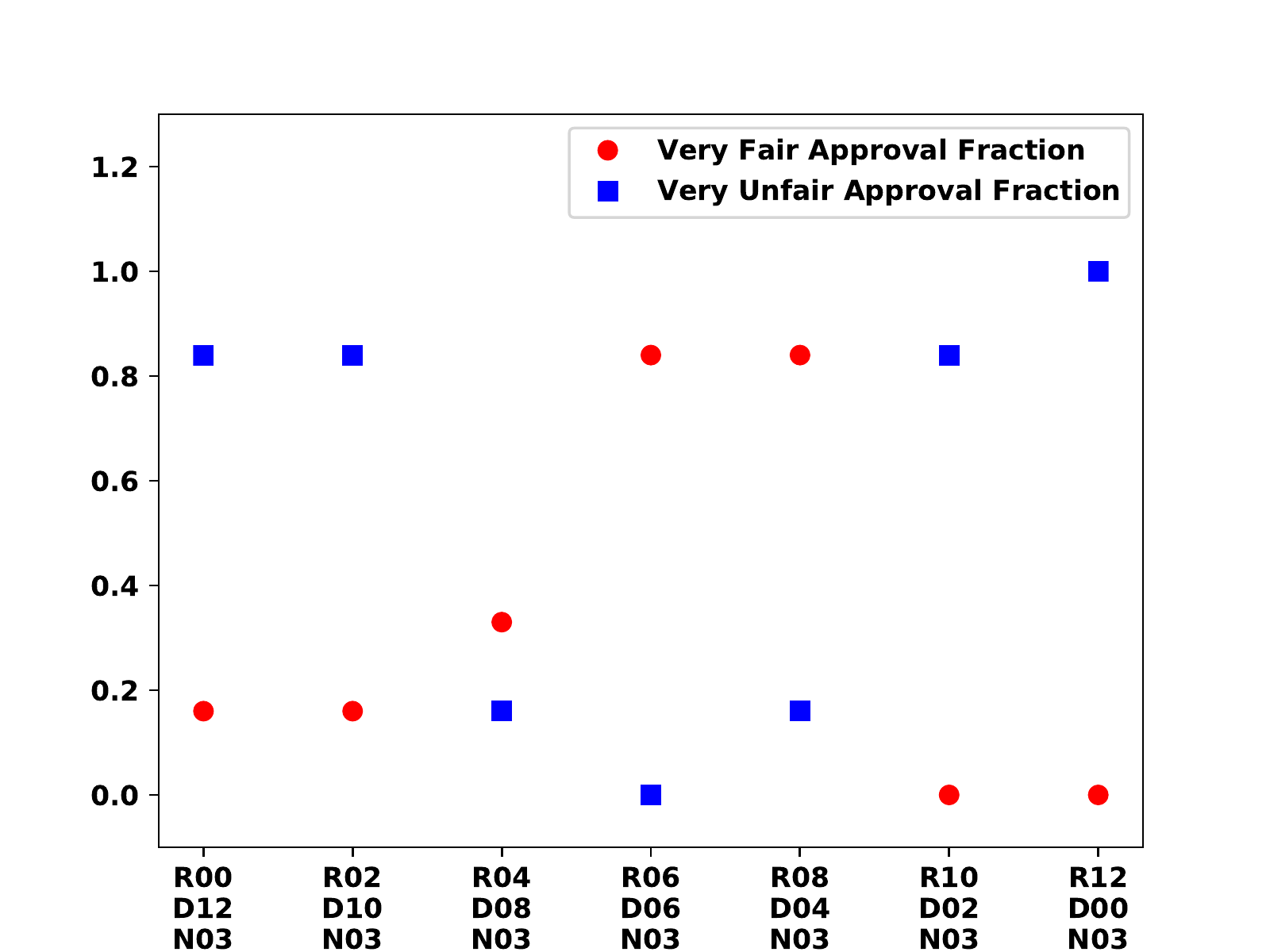}
		\vspace*{-3.5mm}
		\caption{\bf FairSumm-US-Batch1 summaries}
		\label{Fig: AnnotationUSEBatch1}
	\end{subfigure}%
	\hfill
	~\begin{subfigure}{0.65\columnwidth}
		\centering
		\includegraphics[width=\textwidth, height=3cm]{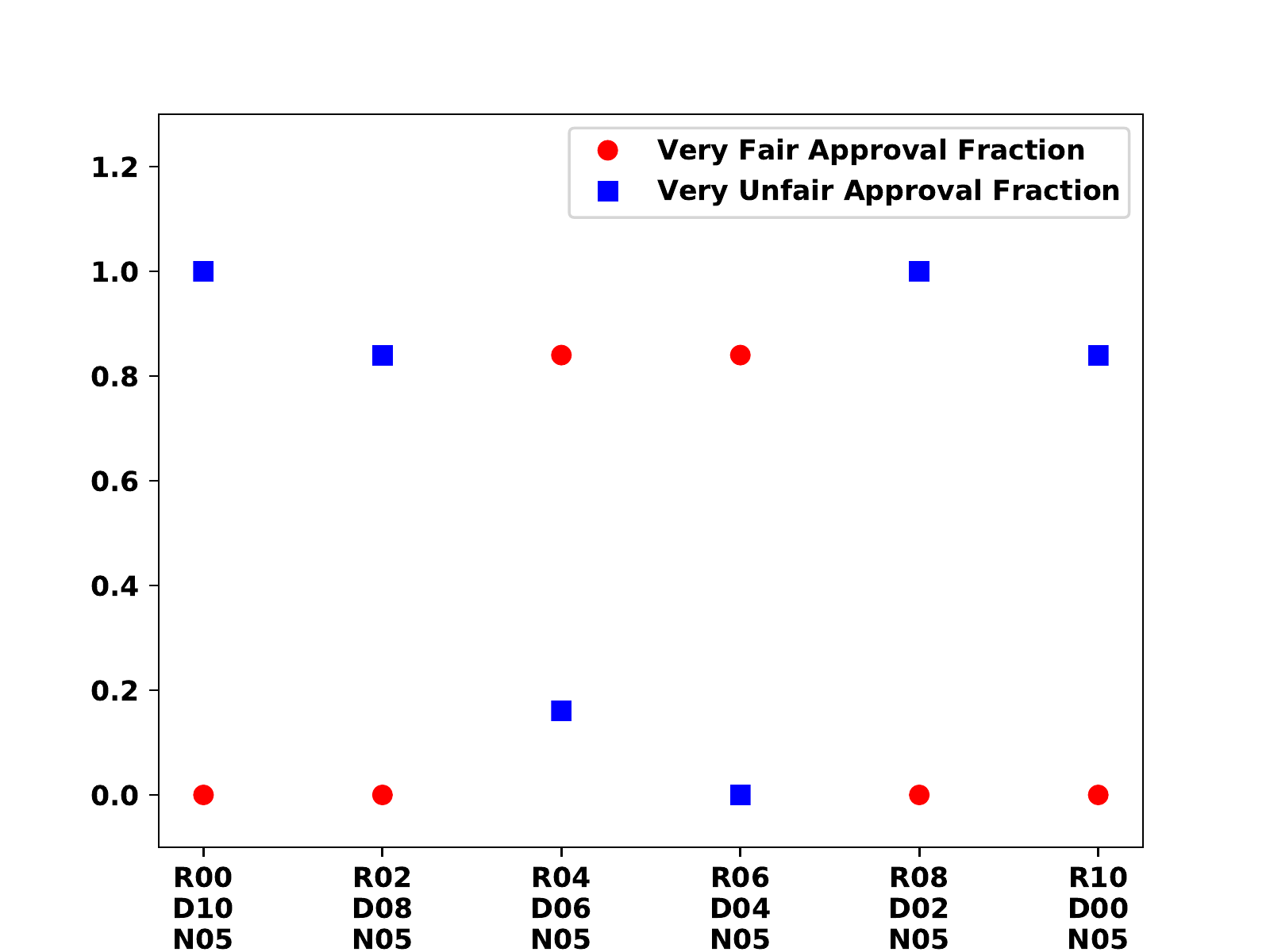}
		\vspace*{-3.5mm}
		\caption{\bf FairSumm-US-Batch2 summaries}
		\label{Fig: AnnotationUSEBatch2}
	\end{subfigure}
	\hfill
	~\begin{subfigure}{0.65\columnwidth}
		\centering
		\includegraphics[width=\textwidth, height=3cm]{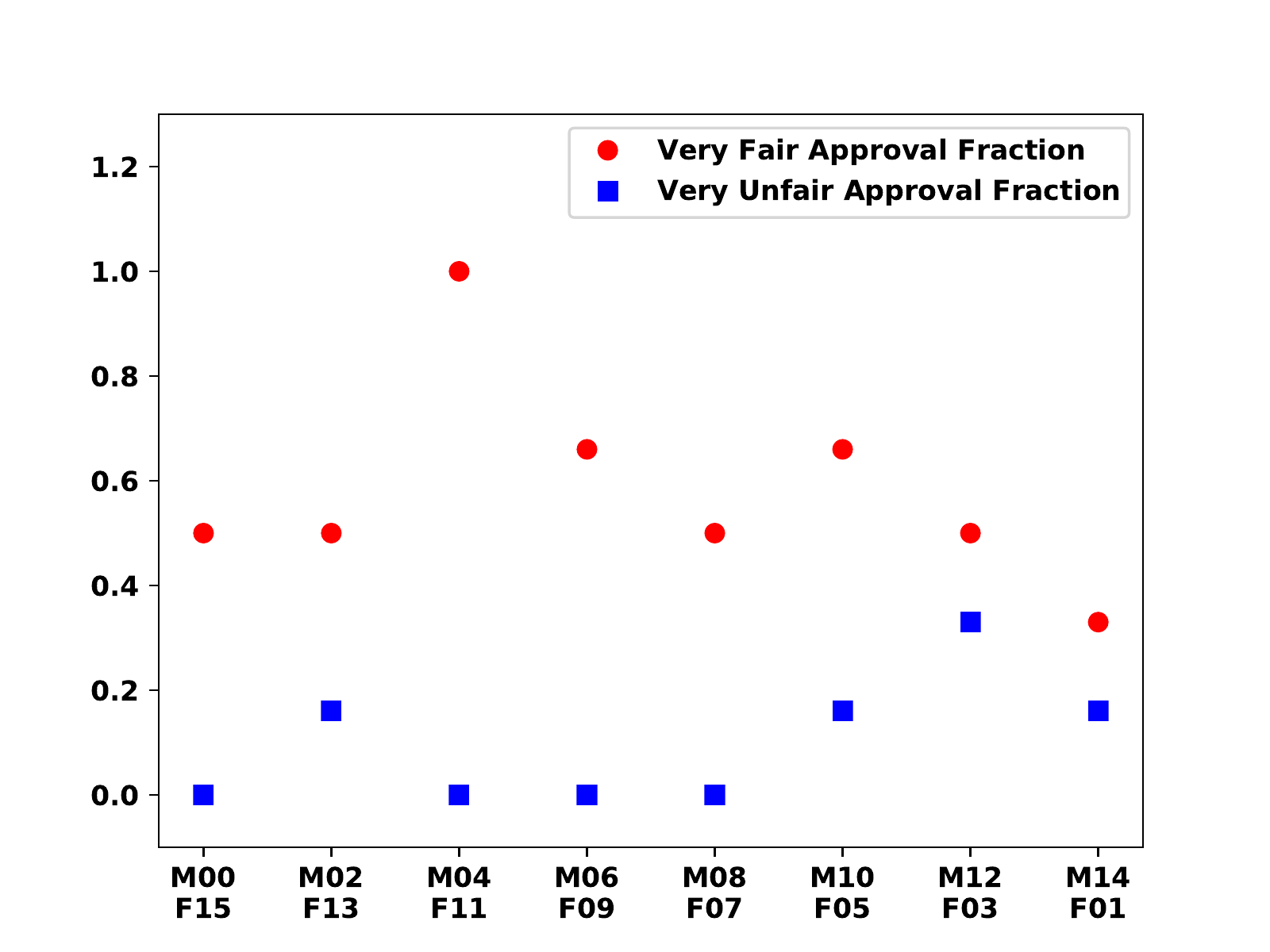}
		\vspace*{-3.5mm}
		\caption{\bf FairSumm-MeToo summaries}
		\label{Fig: AnnotationMeToo}
	\end{subfigure}
	
	\caption{{\bf The fraction of consumers (annotators) who labeled various summaries as `Very Fair Representation' (marked by the red circles) and `Very Unfair Representation' (marked by the blue squares) for USElection dataset with (a)~3 neutral tweets, (b)~5 neutral tweets and (c)~for the MeToo dataset. For the USElection dataset, the majority of consumers' perception of (un)fairness agrees with the actual (un)fairness of the summaries. However, the agreement is much lower for the MeToo dataset.}} 
	\label{Fig: AnnotationDatasets}
	
\end{figure*}

Figure~\ref{Fig: AnnotationDatasets} depicts the result for the USElection dataset (sub-figures (a) and (b) for FairSumm-US-Batch1 and FairSumm-US-Batch1 summaries respectively) and for the MeToo dataset (sub-figure (c) for the FairSumm-MeToo summaries). 
Recall that a batch is a group of summaries having the same number of neutral tweets, but varying number of tweets from other perspectives. 

From the results, it is evident that for the US-Election dataset, the fraction of annotators who said that a summary was `very fair' and the fraction of annotators who said that a summary was `very unfair' correlates well with the actual fairness in the FairSumm summaries. For instance, both summaries having much larger number of pro-Republican tweets and summaries having much larger number of pro-Democratic tweets were labeled as `very unfair' by most annotators. Whereas, the summaries having relatively similar numbers of pro-Republican and pro-Democratic tweets were labeled as `very fair' by most annotators.
Thus, for the USElection dataset, the consumers' perception of fairness in the summaries aligns very well with traditional notions of fairness in representing political groups among the producers (those who authored the tweets).

However, for the MeToo dataset (see Figure~\ref{Fig: AnnotationDatasets}(c)), this is not the case. There is no correlation between group-wise representation of tweets posted by male and female users and the  consumers’ perception of fairness of the summaries. 

This difference for the two datasets leads us to explore more closely why consumers think of a summary as being fair/unfair.

\begin{figure*}[t]
	\centering
	\begin{subfigure}{0.65\columnwidth}
		\centering
		\includegraphics[width=\textwidth, height=4cm]{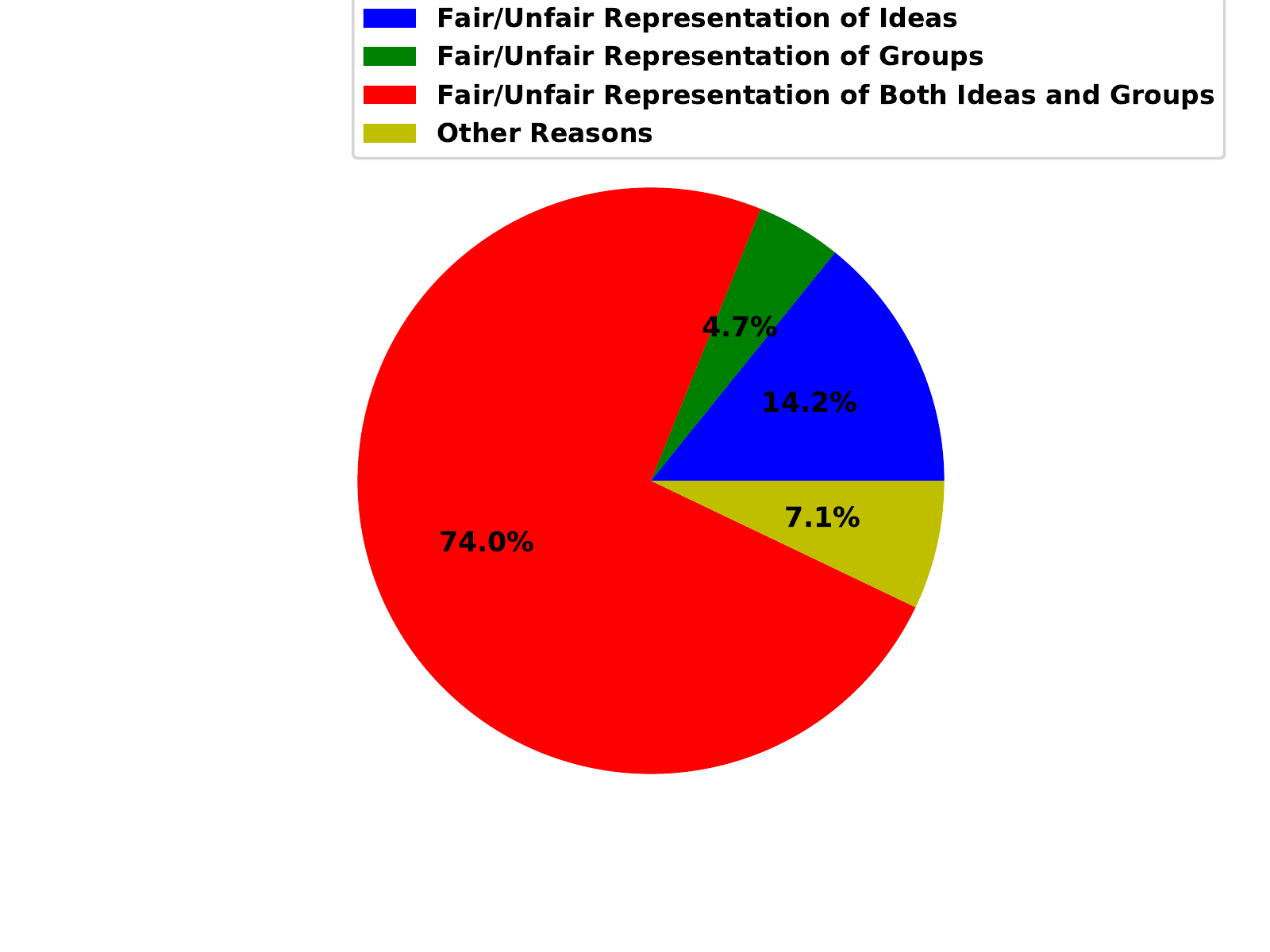}
		\vspace{-10mm}
		\caption{\bf FairSumm-US-Batch1 summaries}
		\label{Fig: ReasonUSEBatch1}
	\end{subfigure}%
	\hfill
	~\begin{subfigure}{0.65\columnwidth}
		\centering
		\includegraphics[width=\textwidth, height=4cm]{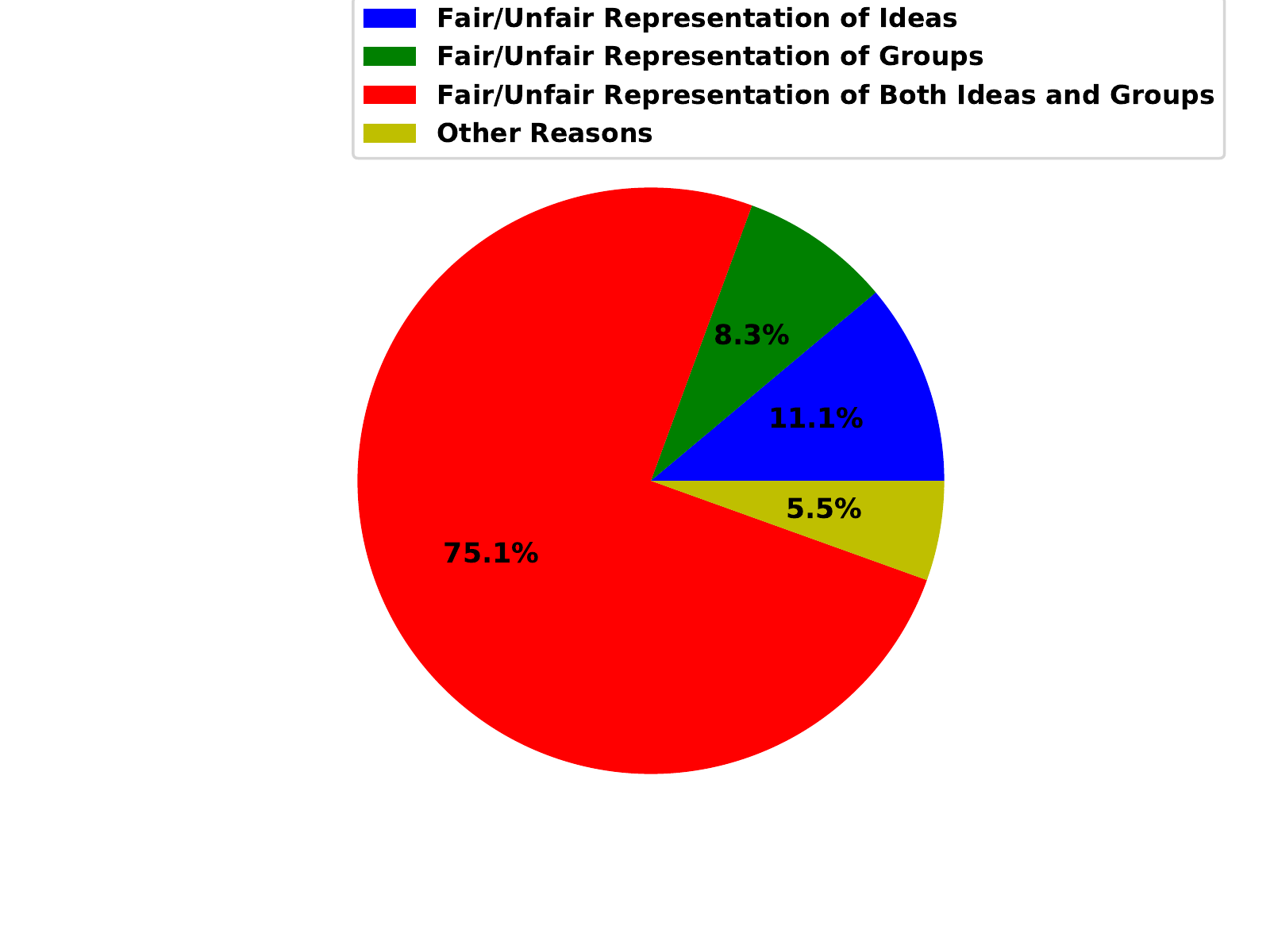}
		\vspace{-10mm}
		\caption{\bf FairSumm-US-Batch2 summaries}
		\label{Fig: ReasonUSEBatch2}
	\end{subfigure}
	\hfill
	~\begin{subfigure}{0.65\columnwidth}
		\centering
		\includegraphics[width=\textwidth, height=4cm]{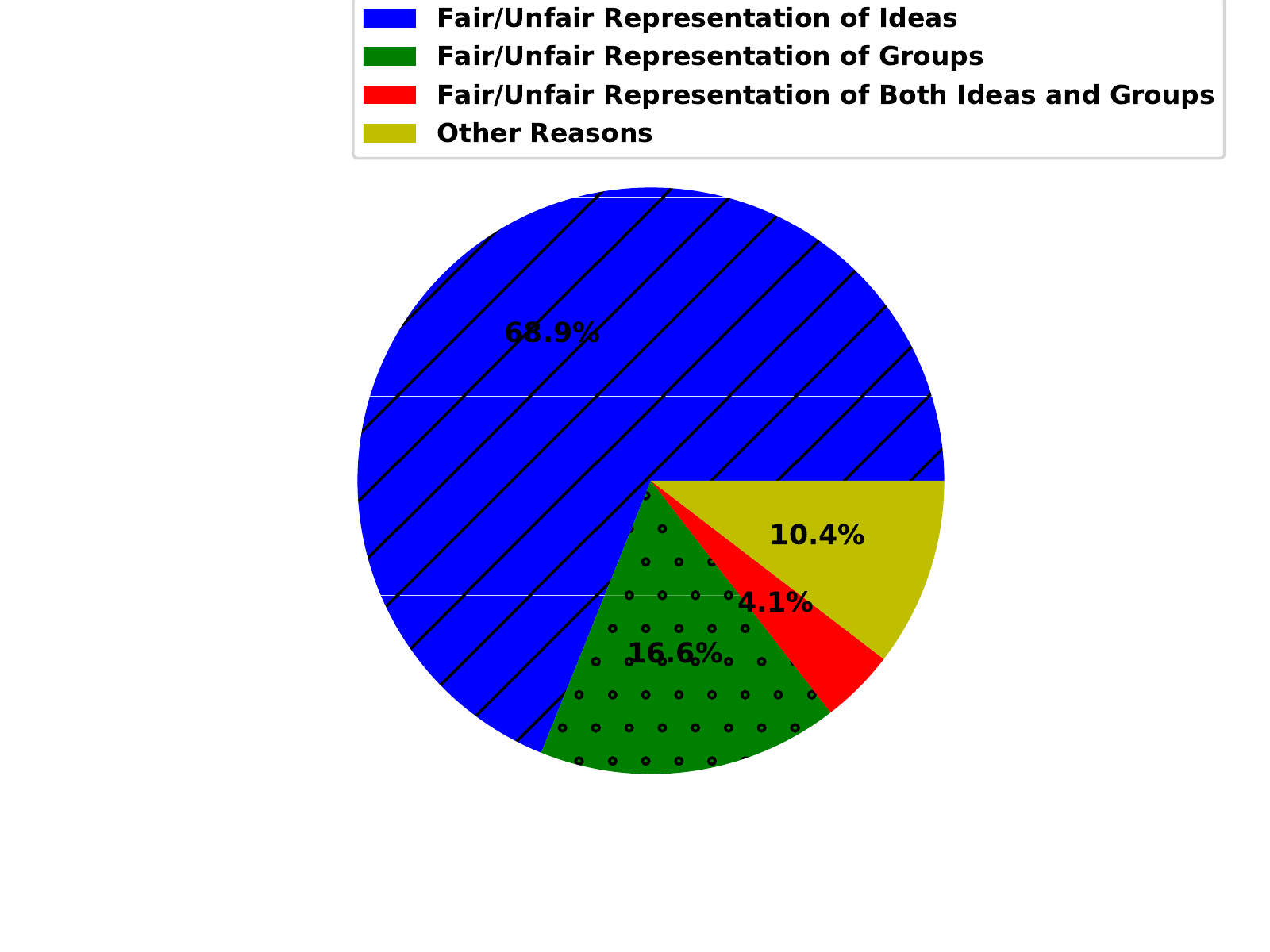}
		\vspace{-10mm}
		\caption{\bf FairSumm-MeToo summaries}
		\label{Fig: ReasonMeToo}
	\end{subfigure}
	
	\caption{{\bf The relative proportions of the various reasons given by annotators for judging a summary as fair/unfair for USElection dataset with (a)~3 neutral tweets, (b)~5 neutral tweets and (c)~for the MeToo dataset. We observe that for the USElection dataset, most consumers labeled a summary to be (un)fair based on the (un)fair representation of both political opinions and groups. Whereas, the consumers gave priority to (un)fair representation of opinions in the MeToo dataset.}} 
	\label{Fig: ReasonDataset}
    \vspace{-5mm}
\end{figure*}

\subsection{Why do consumers think of a summary as being (un)fair?}

\noindent
As stated earlier in this section, we also asked the annotators to indicate {\it why} they labeled a certain summary as fair/unfair -- whether they considered the political/gender groups of the users who posted the tweets (which were not specifically told to them), or the political/contextual opinions (which were identified by the annotators themselves), or both, or some other factor. 
We now look at the distribution of the reasons as stated by the annotators.

Figure~\ref{Fig: ReasonDataset} shows the distribution of reasons, as stated by the annotators, for the three batches of summaries. 
For both batches of the USElection dataset (see Figure~\ref{Fig: ReasonDataset}(a) and Figure~\ref{Fig: ReasonDataset}(b)), the consumers' judgement of fairness/bias is dictated by both the `fair/unfair representation of opinions' and the `fair/unfair representation of political groups’. 
One point to note here is that the consumers (annotators) were {\it not} specifically informed of the group label of the various producers explicitly. However, it is quite evident that they are able to deduce the political group of the author from the textual content of the tweets. One reason for this would be that determination of political grouping is relatively easy if the opinions are properly expressed. 

However, for the MeToo dataset (see Figure~\ref{Fig: ReasonDataset}(c)), 
the situation is different.
In the previous section, it was observed that the consumers’ perception of fairness does {\it not} correlate well with group-wise representation of the producers for this dataset. 
Figure~\ref{Fig: ReasonDataset}(c) gives us an explanation for this observation. 
In the case of the MeToo dataset, the annotators give a disproportionately higher importance to `fair/unfair representation of opinions’ as compared to any other reason. Recall once again the annotators have no knowledge of the groups/class labels (gender in this case) of the producers (those who authored the tweets). 
Thus, it appears that, for this dataset, it was not possible for the consumers to make any inference about group labels from the text of the tweets.

\vspace{3mm}
\noindent {\bf Summary of the section:}
From this section, we have understood that human annotators can understand the fairness/bias of summaries, and their perception of fairness/bias in summaries is dependent on the context of the data. In some cases (e.g., for the USElections dataset), the perceived fairness agrees with standard fairness notions on demographic groups of producers, while in other cases (e.g., for the MeToo dataset), the perceived fairness seems to agree more with how fairly various opinions are represented in the summary.
These results also indicate that proper representation of opinions in the input text is central to the consumers’ idea of fairness in the summaries.

\vspace{3mm}
Next  we  check  whether  traditional  metrics  used for evaluation of summaries  can capture  the  perceived  fairness  of summaries.

\section{Can ROUGE Metrics Capture Consumers' Perception of Fair Summary?} \label{sec:rouge-agreement-with-perception}
\noindent
In the previous section, we have established the important of fair representation of opinions in the consumers’ perception of fairness in summaries. 
In this section, we study the \textbf{RQ$2$} stated in the introduction
-- whether the traditional ROUGE metrics (that are popularly used to measure quality of summaries) can capture the (un)fairness of summaries. 

To this end, we follow the traditional approach of evaluating summaries. 
We first obtain `gold standard' summaries written by human annotators for the two datasets (the set of $50$ tweets related to US Election, and the set of $50$ tweets related to MeToo movement). Then we compute ROUGE scores for the FairSumm-US-Batch1, FairSumm-US-Batch2 and FairSumm-MeToo summaries, considering the gold standard summaries.

\vspace{3mm}
\noindent {\bf Obtaining gold standard summaries for the datasets:}
For the USElection dataset, we conducted a  survey on the Amazon Mechanical Turk (AMT) crowdsourcing platform. 
We selected AMT master workers who are known to be especially skilled in performing data annotation and labeling tasks.
We required that every worker be from the US, and be knowledgeable about US politics. 
We asked them to indicate their political leaning -- Democratic or left-leaning, Republican or right-leaning, or neutral.\footnote{For framing the questions on political leaning, we followed a questionnaire of the Pew Research Center which is a well-known organization for conducting social surveys.} 
We selected $15$ annotators (AMT workers) who are right-leaning and $15$ who were left-leaning, to ensure that we get a balanced set of gold standard summaries. 

During the survey, each AMT worker was shown the $50$ tweets on a screen, and then asked to select the most important $15$ tweets (according to his/her opinion) for generating a summary of the whole set of tweets. Different workers were shown the 50 tweets in different randomly-selected orders, to ensure that the order in which the workers see the tweets do not affect their selection. 

Along the lines of the above survey, we conducted a survey for the MeToo dataset as well. We selected $5$ male annotators and $5$ female annotators for this survey, so that we get a balanced set of gold standard summaries. 
These annotators were shown the $50$ tweets in different randomly-selected orders, and were asked to choose the $15$ most important tweets (according to her/his opinion) for generating a summary of the whole set of tweets. 

\begin{figure*}[t]
	\centering
	\begin{subfigure}{0.65\columnwidth}
		\centering
		\includegraphics[width=\textwidth, height=3cm]{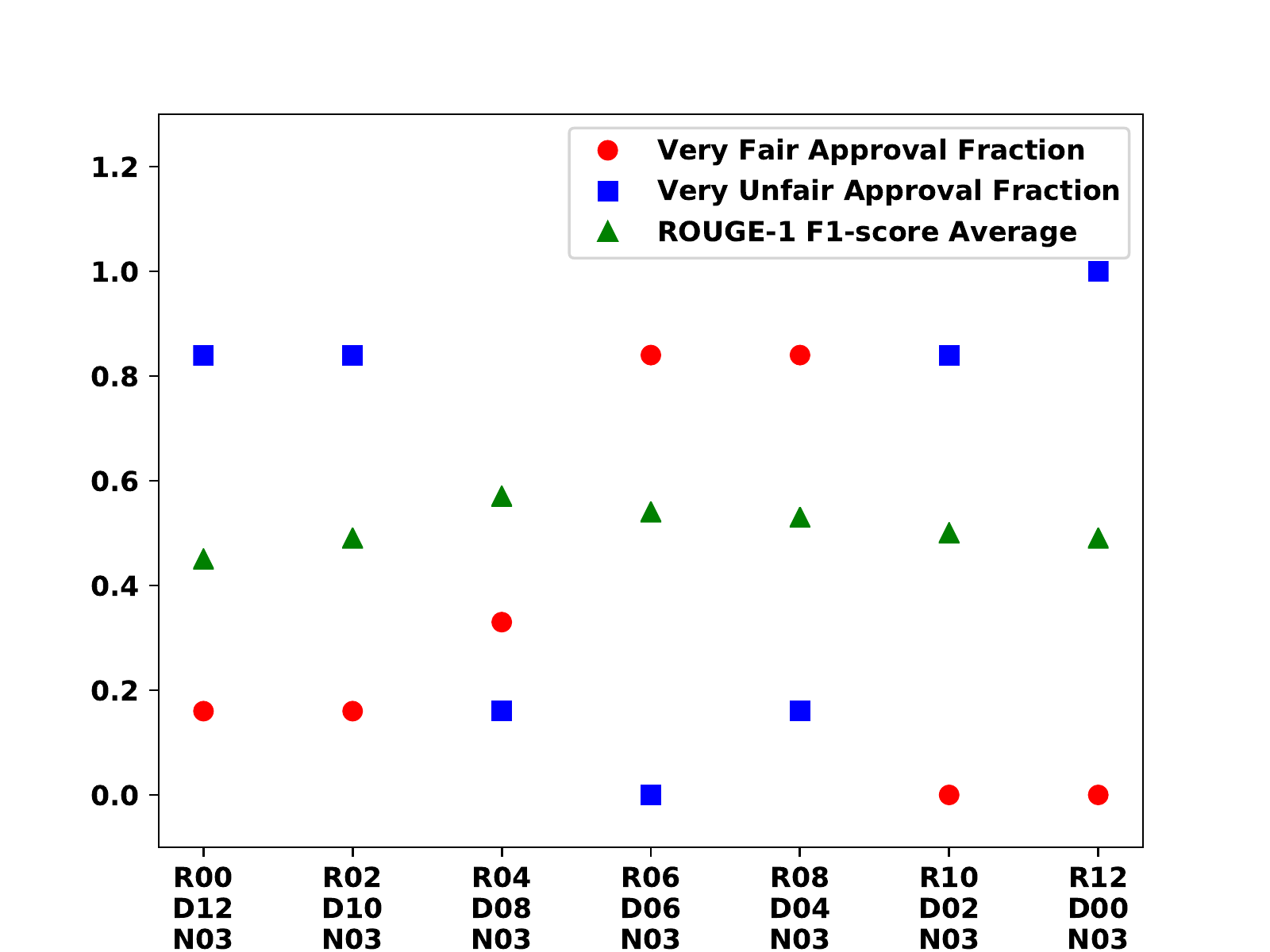}
		\caption{FairSumm-US-Batch1 summaries}
		\label{Fig: RougeApprovalUSEBatch1}
	\end{subfigure}%
	\hfill
	~\begin{subfigure}{0.65\columnwidth}
		\centering
		\includegraphics[width=\textwidth, height=3cm]{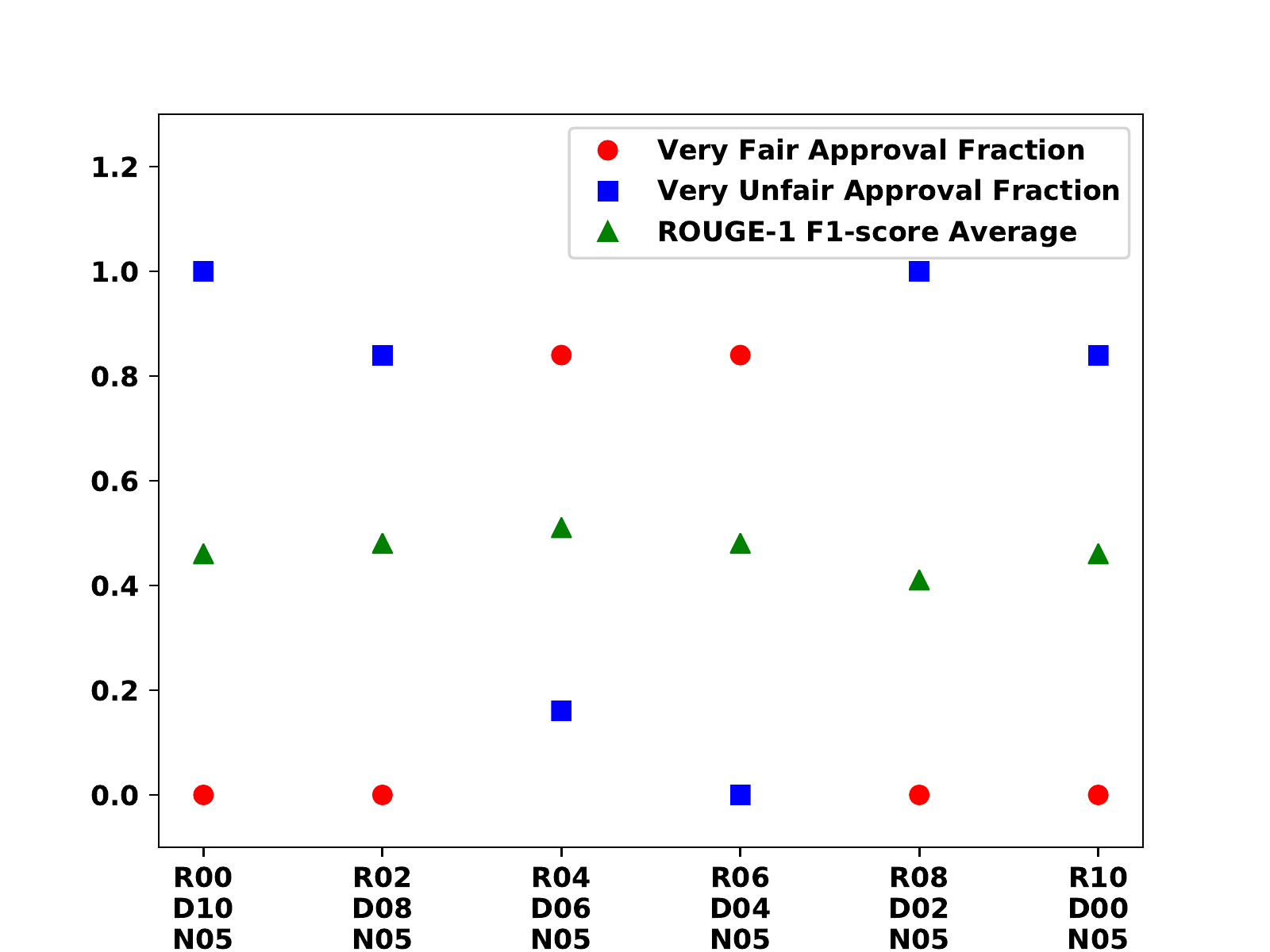}
		\vspace*{-3.5mm}
		\caption{FairSumm-US-Batch2 summaries}
		\label{Fig: RougeApprovalUSEBatch2}
	\end{subfigure}
	\hfill
	~\begin{subfigure}{0.65\columnwidth}
		\centering
		\includegraphics[width=\textwidth, height=3cm]{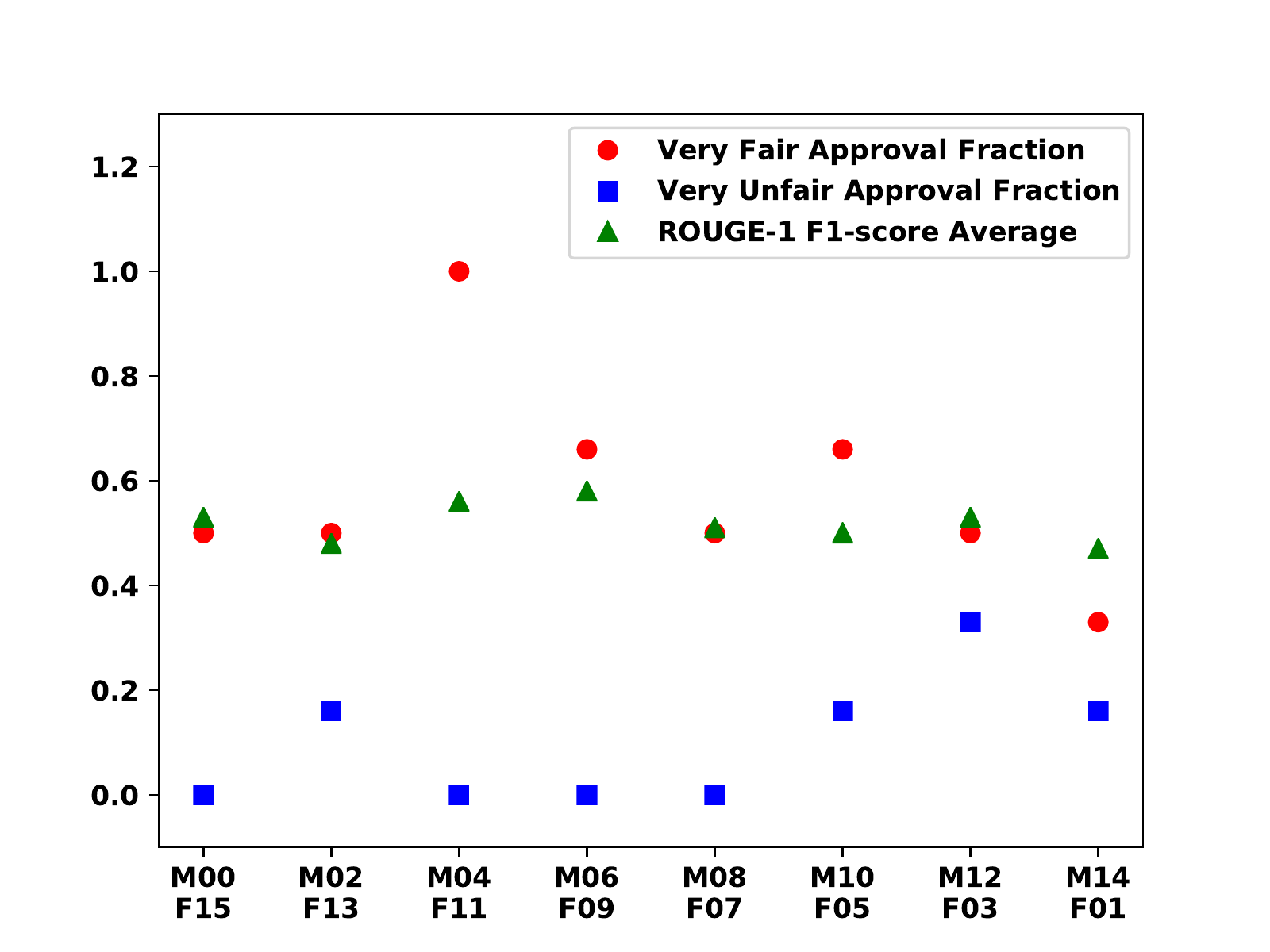}
		\vspace*{-3.5mm}
		\caption{FairSumm-MeToo summaries}
		\label{Fig: RougeApprovalMeToo}
	\end{subfigure}
	\caption{{\bf ROUGE-1 F1-score values for the different summaries (shown by green triangular markers), along with the very fair/unfair approval fractions for the three batches of summaries. In general ROUGE-1 scores have poor correlation with the fairness approval scores, and hence are not a good indicator of fairness of an algorithmic summary.}} 
	\label{Fig: RougeApproval}
\end{figure*}

\vspace{3mm}
\noindent {\bf Computing ROUGE scores:}
We consider the summaries written by the human annotators as `gold standard summaries' and measure the average ROUGE-1 F1 score (based on overlap of unigrams) of all the different summaries in the FairSumm-US-Batch1, FairSumm-US-Batch2 and FairSumm-MeToo batches. 

Note that ROUGE-1 F1 score is computed for an algorithmic summary individually with every gold standard summary (written by a human annotator), and then the average score across all gold standard summaries is considered -- this is in accordance with the standard procedure for evaluation of summaries.

\vspace{3mm}
\noindent {\bf Agreement of ROUGE scores with consumers' perception of fairness in summaries:}
Figure~\ref{Fig: RougeApproval} shows the average ROUGE-1 F1 scores (shown by green triangular markers) obtained by the different summaries in the FairSumm-US-Batch1, FairSumm-US-Batch2 and FairSumm-MeToo batches, along with the fraction of annotators who judged the corresponding summaries to be very fair/unfair (as was described in Section~\ref{sec:consumer-perception}).
Visually, the ROUGE scores appear to have low correlation with the consumers’ perception of fairness of the summaries. 
Very unfair/biased summaries are seen to get similar ROUGE scores as very fair/unbiased summaries. For instance, in Figure~\ref{Fig: RougeApproval}(a), a very biased/unfair summary (containing 12 pro-Republican tweets, 0 pro-Democratic tweets and 3 neutral tweets) obtained a very similar ROUGE score as a very fair summary (containing 6 pro-Republican tweets, 6 pro-Democratic tweets and 3 neutral tweets).

\begin{table*}[tb]
    \centering
    \begin{tabular}{|c|c|c|c|}
    \hline	
     Correlation between   &  FairSumm-US-Batch1 & FairSumm-US-Batch2 & FairSumm-MeToo\\
       \hline \hline
   
     Rouge-1 F1-score \& Very Fair Approval Fraction & 0.51  & 0.61 & 0.64\\
     \hline
     Perceived Opinion Bias Scores \& Very Unfair Approval Fraction & 0.84 & 0.74 & 0.94\\
     \hline
     Perceived Opinion Bias Scores \& Opinion Interaction Graph scores & 0.96 & 0.94 & 0.93\\
     \hline
    \end{tabular}
    \caption{\bf Pearson's correlation coefficient between different metrics/scores as measured for the three batches of summaries. While the ROUGE1 scores do {\it not} correlate strongly with the fairness approval fractions (fraction of annotators who judged a summary to be fair), the proposed metric (Perceived Opinion Bias Score) has a much stronger correlation with the bias/unfairness of summaries as judged by the annotators.}
    \label{tab:correlation-values}
\end{table*}

To quantify the agreement of ROUGE scores with consumers' perception of fairness in summaries, we compute the {\it Pearson correlation coefficient} between the average ROUGE-1 F1-score of a summary and the `very fair approval fraction' (the fraction of annotators who judged the summary to be very fair).
The Pearson correlation coefficients for the three batches of summaries are shown in Table~\ref{tab:correlation-values} (first row).
We observe the Pearson correlation coefficients to be moderate, in the range $[0.5, 0.65]$, for all three batches.

These results show that the popular ROUGE metrics do {\it not} correlate well with the fairness of summaries as perceived by the consumers.

\section{Metric for Capturing Consumers' Perception of Opinion Bias}
\label{sec:bias-metric}

\noindent
Having established that the popular ROUGE scores cannot capture the bias/unfairness in summaries, we now formulate a metric that can capture the bias of summaries with respect to representation of various opinions in the input, as perceived by human annotators. 
In other words, in this section, we study \textbf{RQ$3$} as mentioned in the introduction.

In brief, our proposed bias metric is based on first asking human annotators to identify the set of distinct opinions in the given input text (which is to be summarized), and then checking how well the different opinions are represented in a particular summary. We describe the setup below in detail.

We go back to the survey described in Section~\ref{sec:consumer-perception} where a set of $N=6$ annotators (say, $A_1, A_2, \ldots, A_N$) were asked to identify all the distinct opinions being conveyed by an input set of tweets.
We consider the union of all distinct opinions identified by all the annotators. 
Let the set of all distinct opinions (in the input text) be denoted by $O$, and assume that there are $k$ distinct opinions $O_1, O_2, ..., O_k$.


In an extension of that survey, the annotators were shown the set $O$ of all distinct opinions, and all the summaries (from the batches FairSumm-US-Batch1, FairSumm-US-Batch2 and FairSumm-MeToo) in random order.
For each summary, all the $N=6$ annotators were asked to label whether the summary adequately represents each of the distinct opinions.
More formally, with respect to a particular summary $S$ (that is to be evaluated), we ask each annotator $A_i$ to label each opinion $O_j$ as one of the following, based on which the function $G_{ij}$ is defined as follows -- 
\begin{enumerate}
    \item The opinion $O_j$ is completely represented in the summary $S$ $ \implies G_{ij}(S) = 1.0$
    \item The opinion $O_j$ is somewhat adequately represented in the summary $S$ $ \implies G_{ij}(S) = 0.5$
    \item The opinion $O_j$ is inadequately represented in the summary $S$ $ \implies G_{ij}(S) = -0.5$
    \item The opinion $O_j$ is completely absent in the summary $S$ $ \implies G_{ij}(S) = -1.0$
\end{enumerate}

\begin{figure*}[h]
	\centering
	\begin{subfigure}{0.65\columnwidth}
		\centering
		\includegraphics[width=\textwidth, height=3cm]{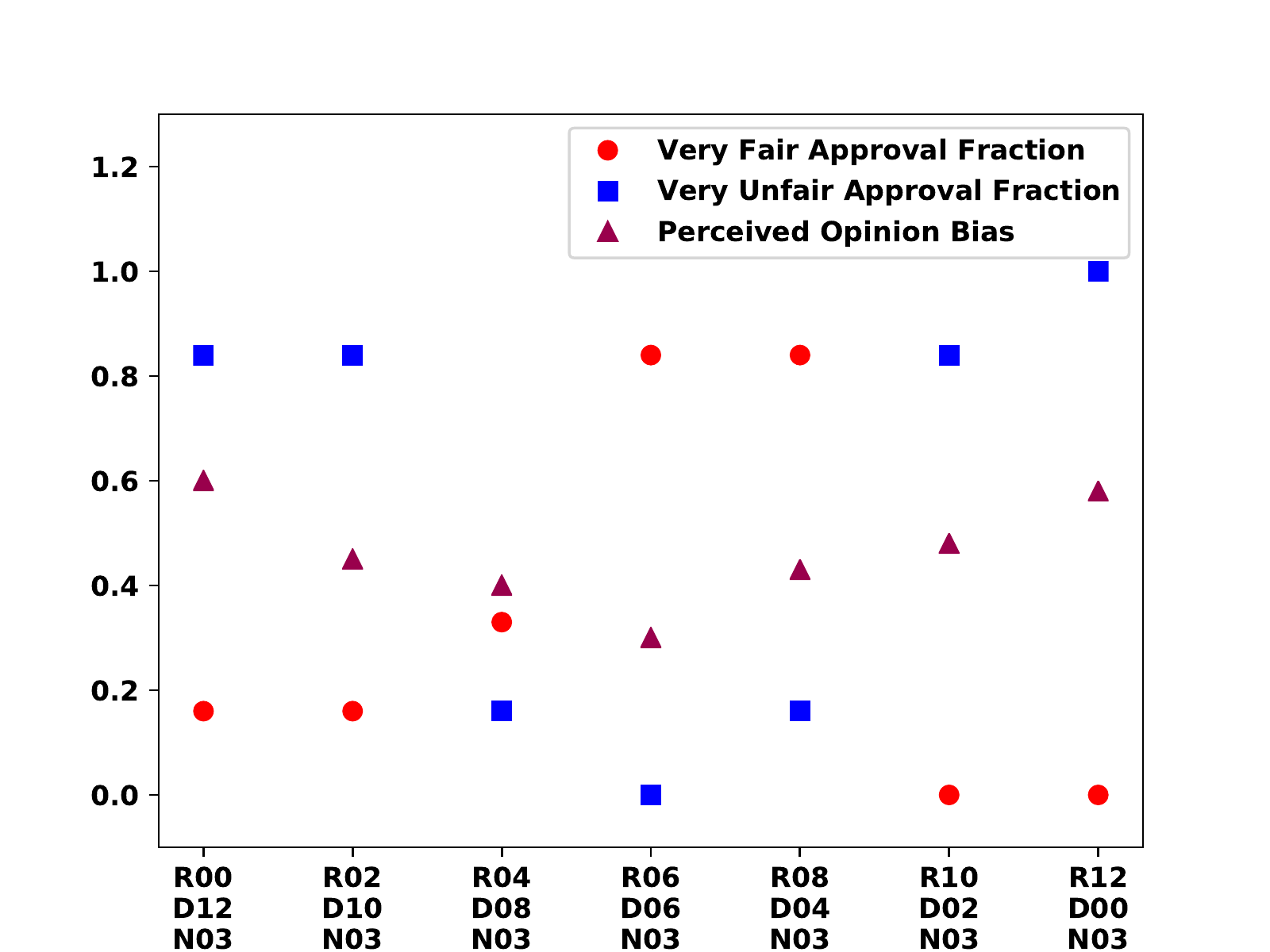}
		\caption{FairSumm-US-Batch1 summaries}
		\label{Fig: HumanApprovalUSEBatch1}
	\end{subfigure}%
	\hfill
	~\begin{subfigure}{0.65\columnwidth}
		\centering
		\includegraphics[width=\textwidth, height=3cm]{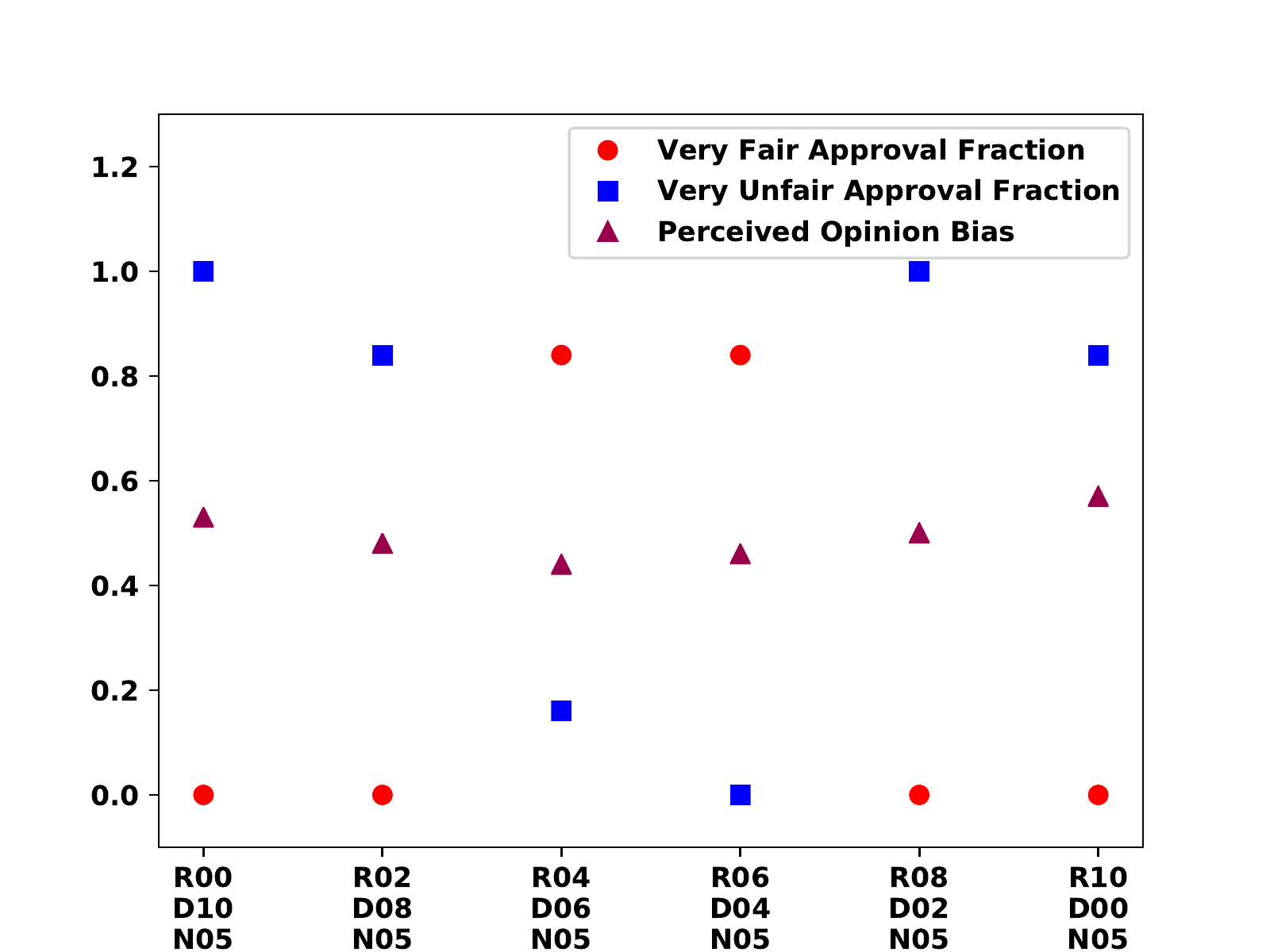}
		\caption{FairSumm-US-Batch2 summaries}
		\label{Fig: HumanApprovalUSEBatch2}
	\end{subfigure}
	\hfill
	~\begin{subfigure}{0.65\columnwidth}
		\centering
		\includegraphics[width=\textwidth, height=3cm]{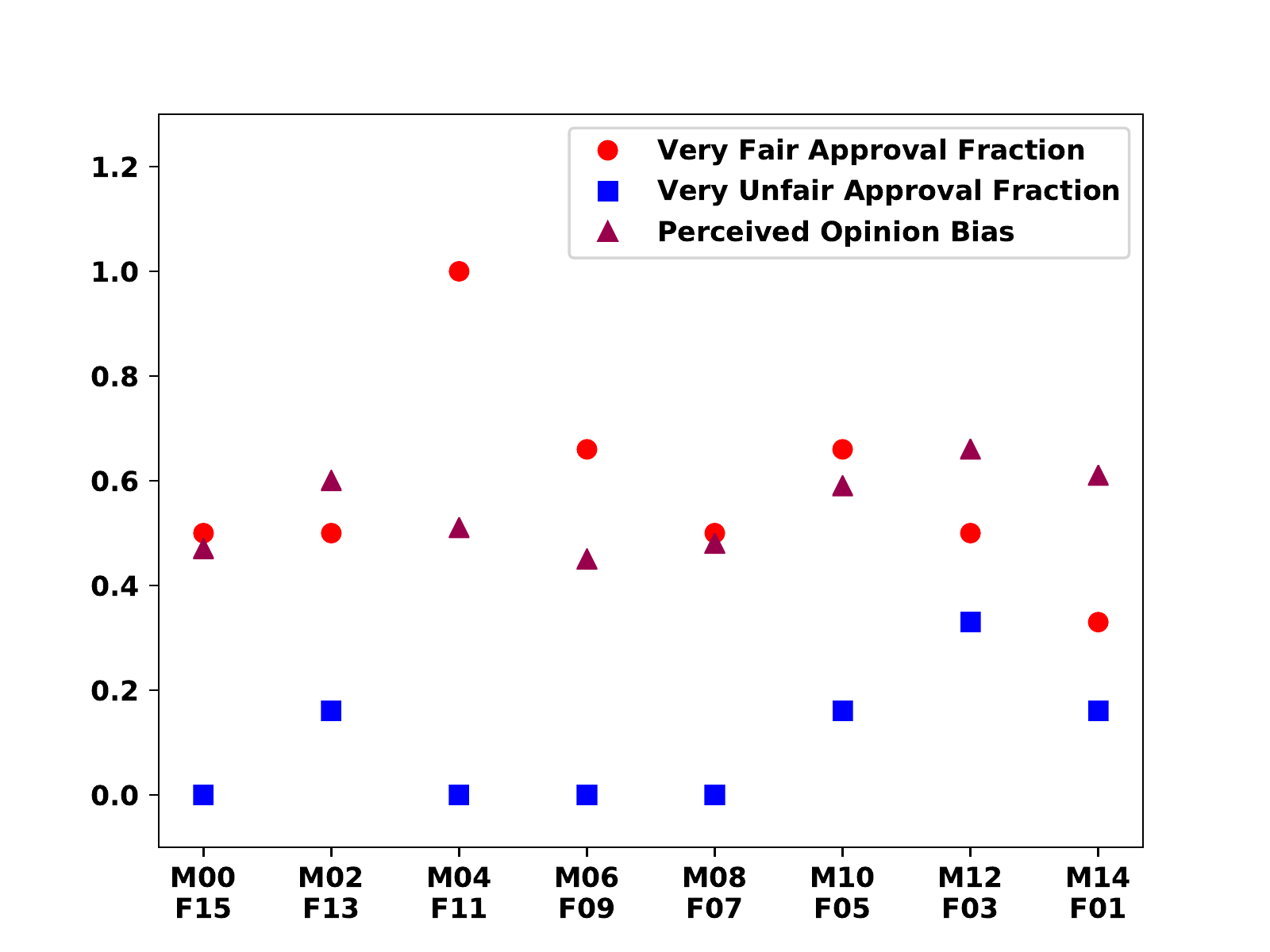}
		\caption{FairSumm-MeToo summaries}
		\label{Fig:HumanApprovalMeToo}
	\end{subfigure}
	\caption{{\bf Perceived Opinion Bias scores (shown by the triangle markers) along with the approval fractions for the three batches of summaries. The Perceived Opinion Bias scores have good agreement with the `very unfair approval fractions'. In other words, the summaries that are judged to be unfair by a high (respectively, low) fraction of annotators have high (respectively, low) Perceived Opinion Bias scores.}} 
	\label{Fig:perceived-opinion-bias}
	\vspace*{-3mm}
\end{figure*}

We define the {\it cumulative representation score} $C_j$  obtained by the opinion $O_j$ in summary $S$ as the mean of all the $G_{ij}$ scores given by all the annotators:
\begin{equation}
   C_j(S) = \frac{1}{N} \sum_{i=1}^N G_{ij}(S)
\end{equation}
Intuitively, $C_j(S)$ denotes how well the opinion $O_j$ is represented in the summary $S$, as judged by all the annotators.

Finally we define the {\it Perceived Opinion Bias} of summary $S$ as the Gini coefficient of the cumulative representation score of all the distinct opinions.
So for a given summary $S$, the perceived opinion bias of $S$ is computed as Gini Coefficient($C_1(S)$, $C_2(S)$,..,$C_k(S)$).
The motivation for using the Gini coefficient is as follows.
The Gini coefficient has been originally used to measure the income inequality or wealth inequality within a group of people (e.g., the people in a certain country). 
Here we apply the Gini coefficient to measure the inequality of representation/exposure within the set of distinct opinions.
If different opinions get widely different amounts of representation/exposure in a summary $S$, then $S$ is biased towards some of the opinions, and hence the Perceived Opinion Bias score of $S$ will be high.

\vspace{3mm}
\noindent {\bf Agreement of Perceived Opinion Bias scores with consumers' perception of unfairness:}
Now we investigate whether our proposed Perceived Opinion Bias scores agree with the consumers' perception of bias/unfairness of summaries. 
Figure~\ref{Fig:perceived-opinion-bias} depicts the perceived opinion bias scores and the very fair/unfair approval fractions (see Section~\ref{sec:consumer-perception} for the definition of these fractions) for the three batches of summaries. 
From the plots, it is evident that there is a good agreement between the perceived opinion bias and the `very unfair approval fraction'.
In other words, those summaries that are judged to be {\it very unfair} by a large fraction of annotators get high Perceived Opinion Bias scores.
In contrast, those summaries that are judged to be {\it very fair} by a large fraction of annotators get low Perceived Opinion Bias scores.

To quantify the agreement, we also compute the Pearson correlation coefficient between the Perceived Opinion Bias score of a summary and the `very unfair approval fraction' (the fraction of annotators who judged the summary to be very unfair).
The Pearson correlation coefficients for the three batches of summaries are shown in Table~\ref{tab:correlation-values} (second row).
For every batch of summaries, we observe the Pearson correlation coefficients to be substantially higher than the corresponding correlation coefficients for the ROUGE F1-scores.\footnote{Note that ROUGE scores are supposed to be higher for good summaries; hence we measure correlation with `very fair approval fraction. In contrast, the Perceived Opinion Bias scores are supposed to be higher for biased/unfair summaries; hence we measure correlation with `very unfair approval fraction'.}
These results show that the proposed Perceived Opinion Bias scores can be used as more reliable measures of bias/unfairness in summaries, than the ROUGE scores.

While the utility of the Perceived Opinion Bias scores is clear, a lot of human annotation effort is needed in computing these scores (first identifying the distinct opinions, and then judging the representation of each opinion in the summary).
Hence the approach of directly computing Perceived Opinion Bias scores may not be scalable to really large datasets.
In the next section, we attempt to develop an {\it automated} methodology for computing the bias/unfairness of summaries.

\section{An Automated Approach to Quantify \\ Consumers' Perceived Bias in Summaries}
\noindent
In this section, we present an automated method to compute the bias/unfairness of summaries.
Our proposed method, inspired by~\cite{concept-interaction-graph}, represents the input text (a document $d$) as an undirected and weighted network/graph, called the {\it Opinion Interaction Graph} ($OIG$). 
We now describe the various steps of the algorithm in detail.

\subsection{The algorithm}

\vspace{2mm} \noindent 
\textbf{Step 1: Generation of Key graph:}
Given the input text, we first extract the named entities and keywords by the TextRank algorithm.
Then we construct a {\it keyword co-occurrence graph}, called {\bf KeyGraph}, based on the set of extracted keywords. Each keyword is a vertex in the KeyGraph. 
We connect two keywords by an edge if they co-occur in the same sentence. The edge between two keywords is weighted by frequency of co-occurrences of the two said keywords.

\begin{figure*}[t]
	\centering
	\begin{subfigure}{0.65\columnwidth}
		\centering
\includegraphics[width=\textwidth, height=3cm]{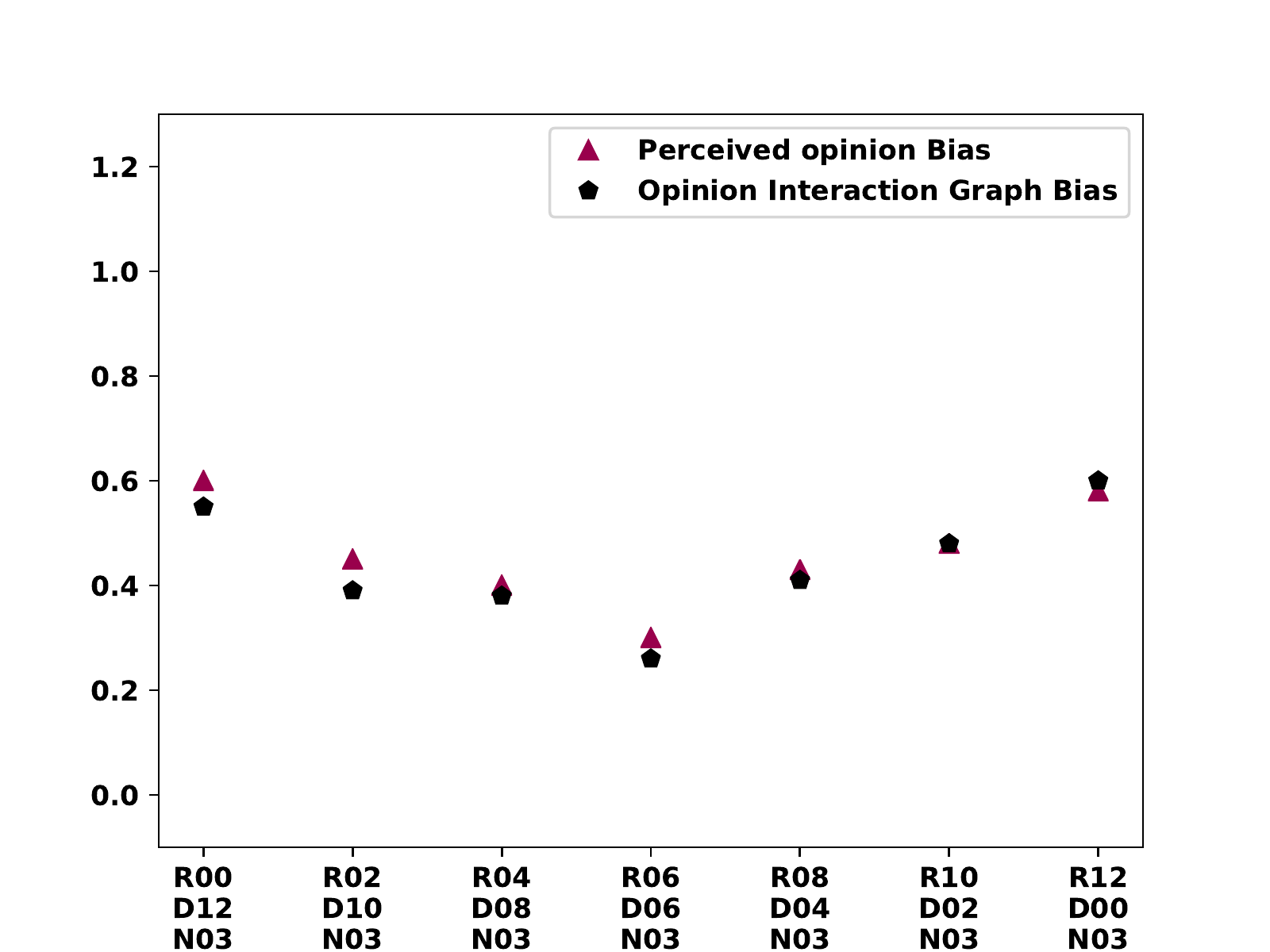}
		\caption{FairSumm-US-Batch1 summaries}
		\label{Fig: GraphBiasScoreUSEBatch1}
	\end{subfigure}%
	\hfill
	~\begin{subfigure}{0.65\columnwidth}
		\centering
		\includegraphics[width=\textwidth, height=3cm]{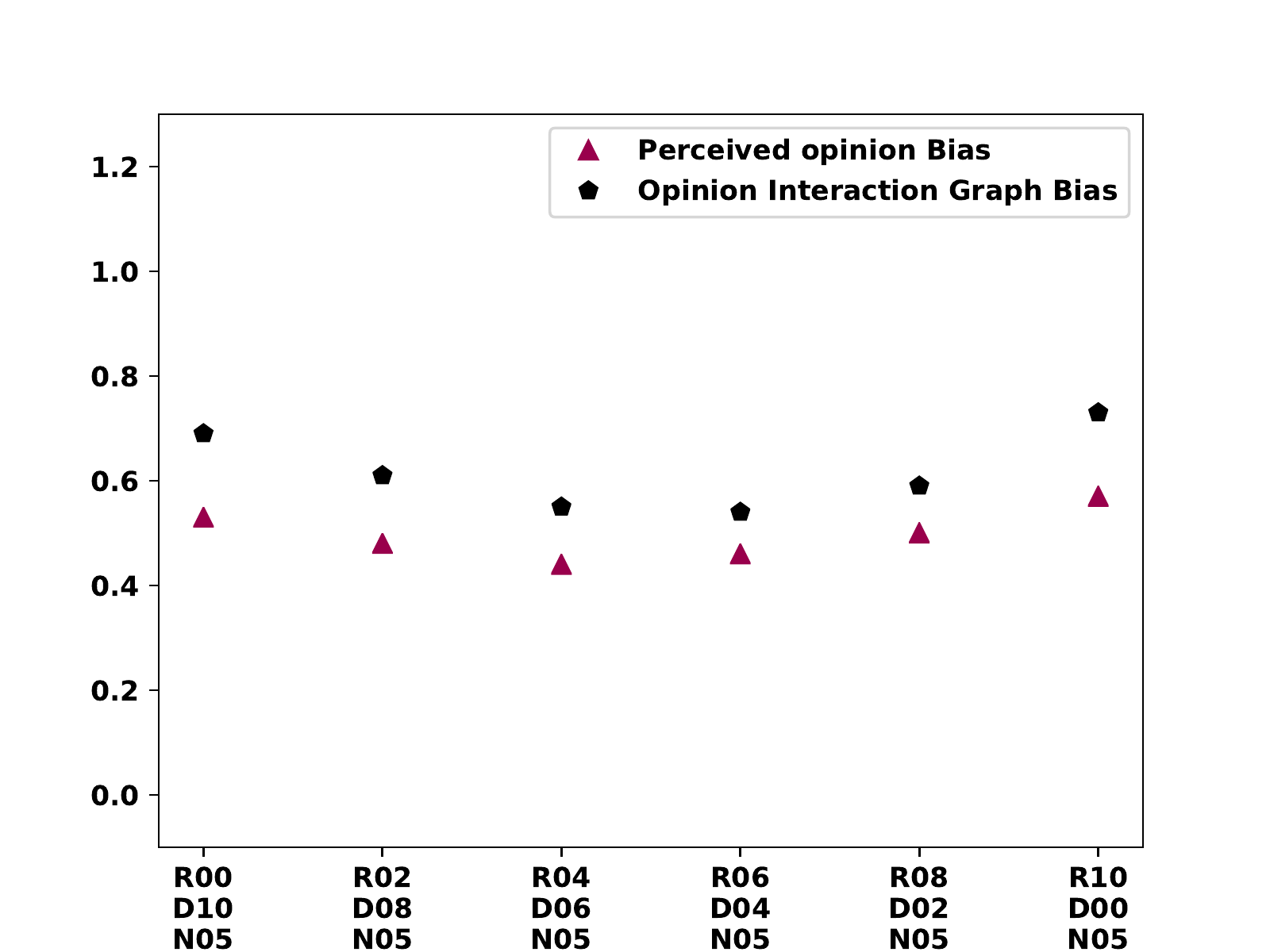}
		\caption{FairSumm-US-Batch2 summaries}
		\label{Fig: GraphBiasScoreUSEBatch2}
	\end{subfigure}
	\hfill
	~\begin{subfigure}{0.65\columnwidth}
		\centering
		\includegraphics[width=\textwidth, height=3cm]{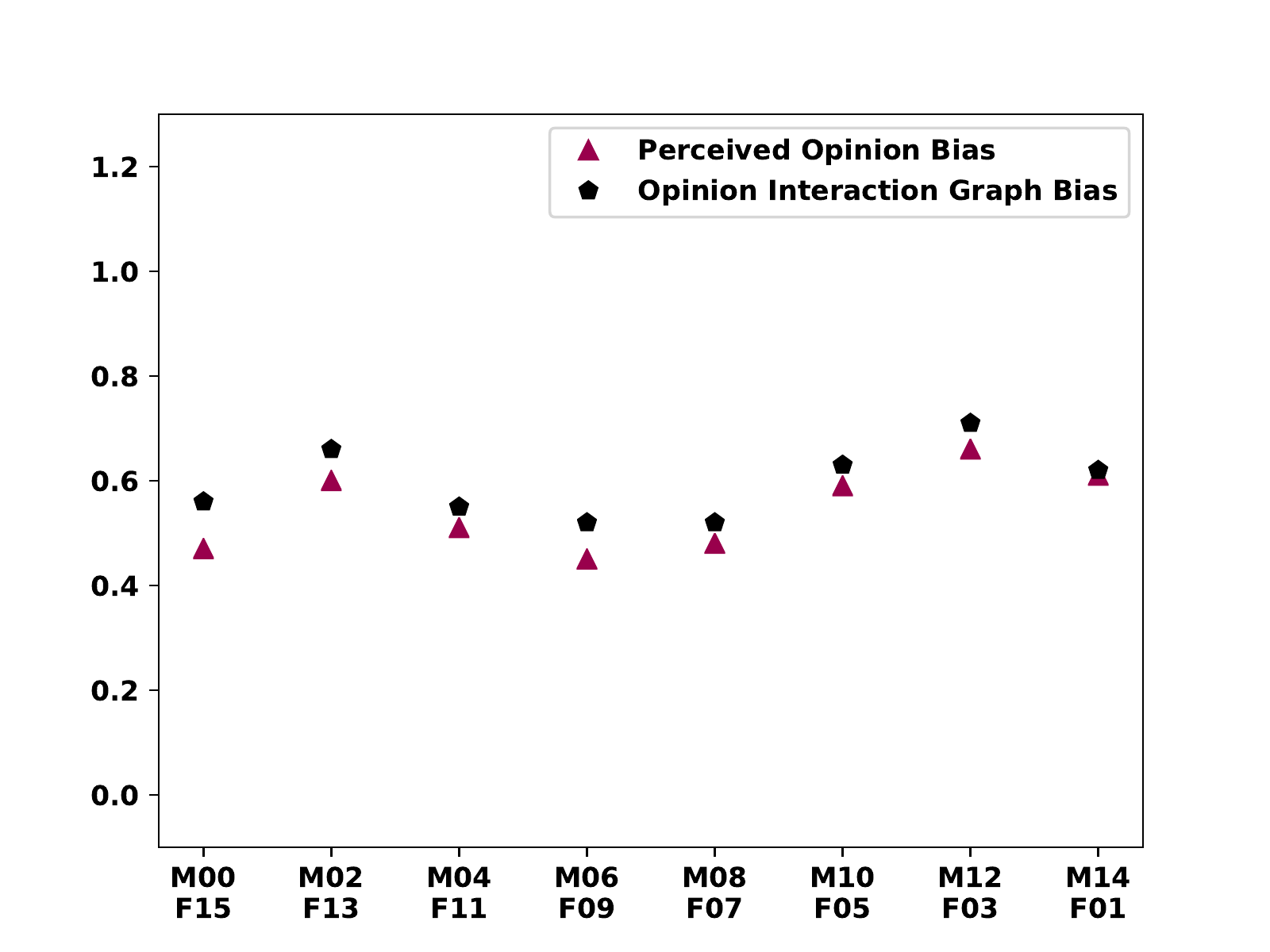}
		\caption{FairSumm-MeToo summaries}
		\label{Fig: GraphBiasScoreMeToo}
	\end{subfigure}
	
	\caption{{\bf Opinion Interaction Graph scores (computed automatically) along with Perceived Opinion Bias scores (computed based on human annotation) for the three batches of summaries. The two scores have very high agreement, thus establishing that our methodology based on Opinion Interaction Graph can be used to automatically measure the Perceived Opinion Bias of summaries.}} 
	\label{Fig: GraphBiasScore}
\end{figure*}

\vspace{2mm} \noindent 
\textbf{Step 2: Concept Detection:}
The structure of KeyGraph reveals the connections between keywords. If a subset of keywords are highly correlated, they will form a densely connected subgraph in  KeyGraph, which we call an {\it opinion}.\footnote{Intuitively, the idea is similar to what is followed in topic modeling, where each topic is essentially a set of frequently co-occurring terms.} 
Opinions can be extracted by applying {\it community detection algorithms} on the KeyGraph. 
A Community Detection algorithm is used to split a KeyGraph  into a set of communities $O$ = \{$O_1$, $O_2$, .., $O_{|O|}$\}, where each community $O_i$ contains the keywords related to a certain opinion. 
To this end, we use the popular Louvain community detection algorithm~\cite{blondel-louvain} for clustering the KeyGraph into fixed sized  communities. However, other clustering methods can also be used in this step. 

\vspace{2mm} \noindent 
\textbf{Step 3: Sentence Attachment and Edge Construction:}
After the opinions are discovered, the next step is to associate sentences to opinions. We calculate the cosine similarity between each sentence and each opinion, where sentences and opinions are represented by TF-IDF vectors of the words. 
We assign each sentence to that opinion $O_i$ which is the most similar to the said sentence, where the similarity is computed based on what fraction of the keywords associated with an opinion is contained in the said sentence.
The sentences that do not match any opinion in the document will be attached to a dummy vertex that does not contain any keywords. 

Then we construct the Opinion Interaction Graph $OIG$ where each vertex/node is an opinion.
To construct edges that reveal the similarity between different opinions, for each vertex, we represent its associated set of sentences as a concatenation of the sentences attached to it. The edge weight between two vertices is computed as the TF-IDF similarity between their associated sentence sets.

\vspace{2mm} \noindent 
\textbf{Step 4: Computing exposure of an opinion in a summary:}
As of now, we have constructed the $OIG$ where every node is an opinion and is associated with a set of sentences. Next we quantify the representation/exposure of different opinions in a given summary $S$ (which is to be evaluated).
We simply compute the exposure of an opinion as the fraction of the sentences attached to the said opinion, that is present in the summary $S$.\footnote{Note that more complex models can be applied to compute the exposure of opinions, e.g., a part of the exposure of $O_j$ can be thought to diffuse to another very similar opinion $O_j$, where the similarity between the two opinions is quantified by the edge-weight in the $OIG$.
However, we have avoided such complexities in order to keep our model simple.}



\vspace{2mm} \noindent 
\textbf{Step 5: Quantifying the skew in the distribution of exposure:} 
Finally, we compute the Gini coefficient of the exposure received by all the distinct opinions (as computed above) to quantify the bias in the distribution of exposure of different opinions in the summary $S$.
The intuition behind using the Gini coefficient has been discussed in Section~\ref{sec:bias-metric}.

It can be noted that, intuitively, we adhere to the {\it proportional representation notion of fairness} (that was explained in Section~\ref{sub:fairness-notions}) among the exposures obtained by different opinions. 
In other words, a summary would be considered most fair if the distribution of exposure received by the various opinions resembles the distribution of sentences attached to the opinions.

\subsection{Results based on opinion interaction graph}
\noindent
Figure~\ref{Fig: GraphBiasScore}
shows the bias of the various summaries in the three batches, as computed by our proposed Opinion Interaction Graph algorithm, and the Perceived Opinion Bias scores of the summaries as obtained in the previous section. 
It is evident that there is a very high correlation between the metrics.

Also Table~\ref{tab:correlation-values} (last row) shows the Pearson correlation for the Perceived Opinion Bias scores and the bias scores computed by the OIG-based method. For all three batches of summaries, the correlation scores are above $0.9$.

These results show that our proposed graph-based algorithm is a good proxy for automatic calculation of perceived opinion bias of summaries.

\section{Conclusion}
\noindent
To our knowledge, this work is the first attempt to explore fairness in the context of automatic summarization from the perspective of consumers/readers of the summary. 
We show that the notion of fairness in summaries from the consumers' perspective varies from one context to another (e.g., may correspond to fair representation of demographic groups of the producers/writers, or the fair representation of opinions from the input text).
Also the popular ROUGE metrics for evaluation of summaries usually cannot capture the fairness of summaries.
To bridge this gap, we have proposed an alternative metric for measuring the bias in summaries (based on human annotation), as well as an automatic  methodology to approximate the metric.

We believe that this work has several potential applications in areas where the text to be summarized consists of multiple different perspectives or opinions, e.g., in news article summarization, debate summarization, and so on.
We plan to explore such applications in future. 
Also, we plan to develop metrics that can simultaneously capture both the quality and the fairness of summaries, e.g., by suitably combining the ROUGE metrics with the bias metric proposed in this work.

\section*{Acknowledgments}
\noindent
The authors would like to thank the annotators who judged the summaries as part of the work. 
This research was supported in part by a European Research Council (ERC) Advanced Grant for the project ``Foundations for Fair Social Computing'', funded under the EU Horizon 2020 Framework Programme (grant agreement no. 789373). 
A. Dash was supported by a fellowship from Tata Consultancy Services.



%

\bibliographystyle{IEEEtran}
\bibliography{fairness.bib}

\end{document}